\documentclass[reprint, amsart, amsmath,amssymb, aps, superscriptaddress,pdftex]{revtex4-1}
\pdfoutput=1
\usepackage{graphicx}
\usepackage{dcolumn}
\usepackage{bm}
\usepackage{mathtools}
\usepackage{float}

\usepackage{dcolumn}   

\begin{document}
\pagenumbering{arabic}

\title{Spatial interactions and oscillatory tragedies of the commons}

\author{Yu-Hui Lin}
\affiliation{School of Physics, Georgia Institute of Technology, Atlanta, GA,
USA}
\author{Joshua S. Weitz}
\affiliation{School of Physics, Georgia Institute of Technology, Atlanta, GA,
USA}
\affiliation{School of Biological Sciences, Georgia Institute of Technology,
Atlanta, GA, USA}
\date{\today}
\begin{abstract}
A tragedy of the commons (TOC) occurs when individuals 
acting in their own self-interest deplete commonly-held resources, 
leading to a worse outcome than had they cooperated.
Over time, the depletion of resources can
change incentives for subsequent actions.
Here, we investigate
long-term feedback between game and environment 
across a continuum of incentives in an individual-based framework.
We identify payoff-dependent transition rules that lead to 
oscillatory TOC-s in stochastic simulations and the mean field limit.
Further extending the stochastic model, we find that spatially explicit interactions 
can lead to emergent, localized dynamics, 
including the propagation of cooperative wave fronts and cluster formation
of both social context and resources.
These dynamics suggest new mechanisms underlying how TOCs
arise and how they might be averted.
\end{abstract}
\maketitle

In 1968, Garrett Hardin explored a social dilemma, which he termed the `Tragedy
of the Commons' (TOC) \cite{Hardin1243}. The social dilemma arises when
two individuals choose amongst distinct strategies
to utilize a limited public good.
Both individuals receive the maximal combined benefit
if they utilize the public good with restraint, i.e., if they `cooperate'. 
However, each individual receives the maximal personal benefit
if they utilize the public good without restraint, i.e., if they `defect',
while their opponent cooperates.  As a consequence,
individuals acting rationally will cheat leaving all worse off.
Hardin argued that
such a TOC is inevitable~\cite{Hardin1243}.
\begin{figure*}[t!]
\centering
	\includegraphics[width=0.45\textwidth]{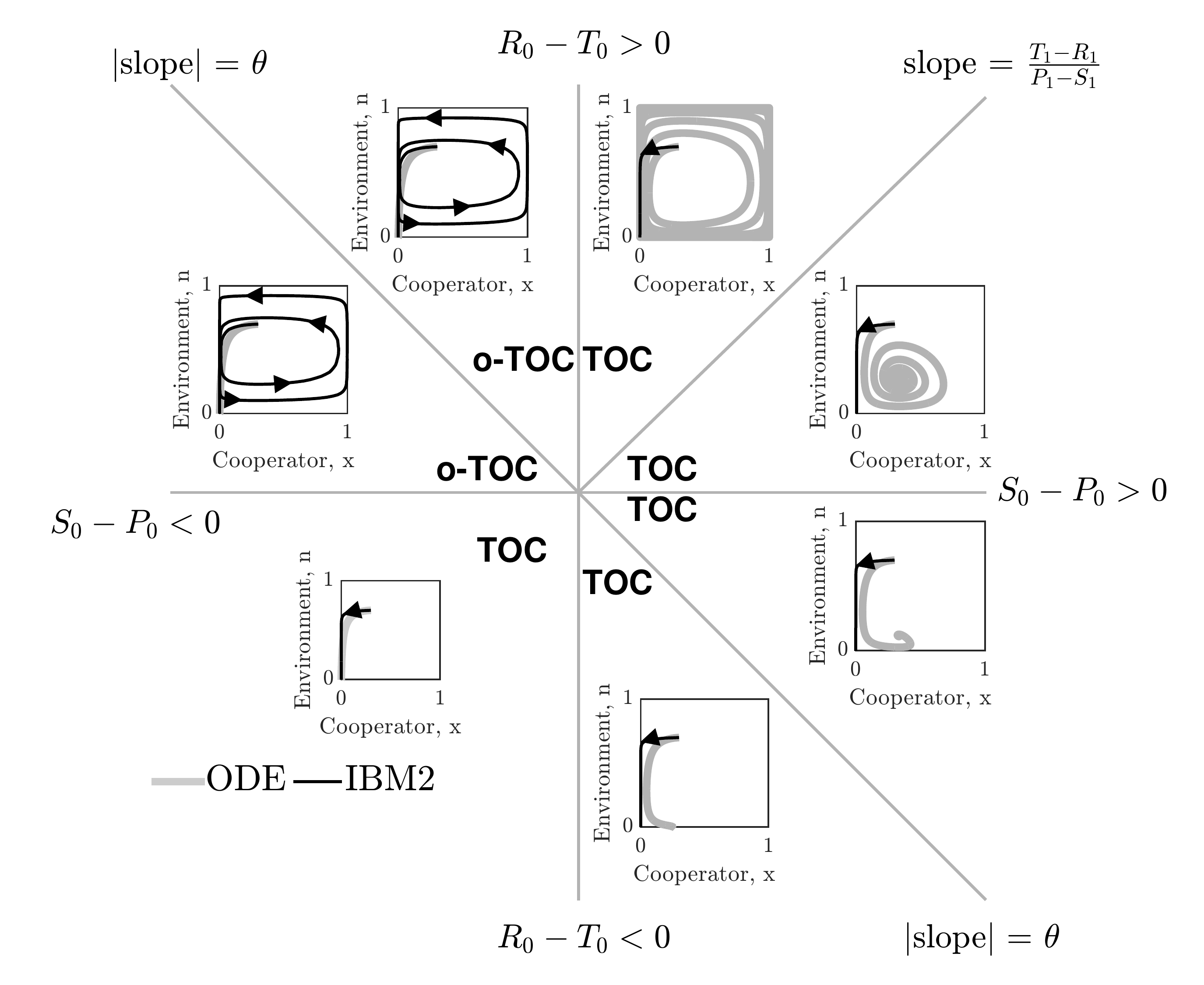}
	\includegraphics[width=0.45\textwidth]{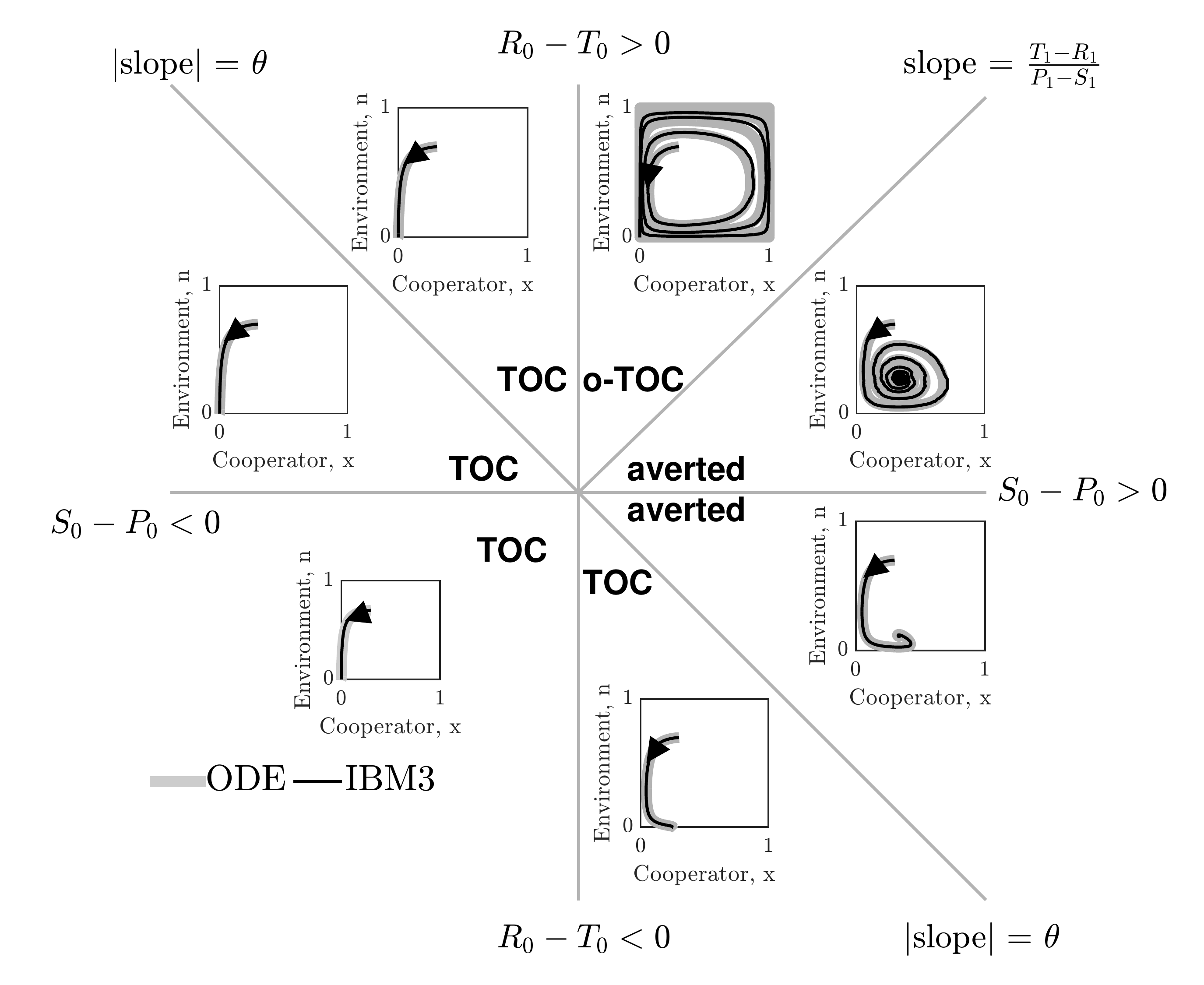}
\caption{\footnotesize Coevolutionary dynamics of strategies and resources in replicator and IBM dynamics. (a) The dynamics with IBM2, in which offspring of the focal player replace the opponent. (b) The dynamics with IBM3, in which offspring of the focal player replaces a random individual.  In both panels, parameter space is divided according to the sign of \(R_0-T_0\), \(S_0-P_0\). In each section in the parameter space, a phase diagram with different $A_0$ in is shown, where the x-axis represents $x$ and the y-axis denotes $n$. Light gray trajectories are mean field solutions and black trajectories denote IBM dynamics where arrows denote the flow of time. Visualized IBM trajectories are the average of 100 replicates with the same parameter set, except for oscillatory dynamics, given phase differences that can arise due to demographic noise. Common parameters for all replicates: $\theta=2$, $\epsilon = 0.5$, $\Delta x =1$, and $\Delta t = 0.05$, $A_1 = [3, 0;5, 1]$; $A_0$ varies by region. Full parameter list for $A_0$
in FIG. S1.}
\label{fig:IBM_ODE}
\end{figure*}

Evolutionary dynamics arising from a TOC dilemma can be modeled 
in terms of changes in the frequencies of individuals from two populations.
Indivdiuals interact and receive payoffs that depend on their
strategy and the strategy of their
opponent. In replicator dynamics~\cite{nowak2006evolutionary},
the payoff represents a relative fitness which
determines the growth of cooperators, with frequency $x$, 
and of defectors, with frequency $1-x$, i.e.,:
\begin{equation}
\label{eq:classical_replicator}
	\dot{x}=x(1-x)(r_C(x,A)-r_D(x,A)).
\end{equation}
The values $r_C$ and $r_D$ denote the context-dependent 
fitness payoff to cooperators and defectors respectively, given the payoff matrix \cite{smith1973logic}, 
$A = \begin{bsmallmatrix} R & S \\T & P\end{bsmallmatrix}$,
where $r_C=Rx+S(1-x)$, $r_D=Tx+P(1-x)$ and
where $R$ denotes the reward to cooperation, $T$ denotes the
temptation to cheat, $S$ denotes the sucker's payoff, and
$P$ denotes the punishment given mutual defection. A TOC occurs 
when $T>R$, $P>S$, and $P<R$. However, in contrast to standard 
game theory assumptions,
payoffs are unlikely to remain fixed after repeated decisions 
that degraded commonly-held resources. 

To address this issue, a recent model \cite{WeitzE7518}
considered dynamics arising given
resource-dependent payoff matrices
$A(n)=A_0(1-n)+A_1(n)$, which interpolate between
$A_0$ and $A_1$, the payoff matrices given deplete and
replete resource states, respectively, \emph{i.e.},
$A(n)=\begin{bsmallmatrix}
		R_0 & S_0 \\
		T_0 & P_0
	\end{bsmallmatrix}
(1-n)+
	\begin{bsmallmatrix}
		R_1 & S_1 \\
		T_1 & P_1
	\end{bsmallmatrix} 
n
$.
This model of coevolutionary game dynamics
included feedback with the environmental state denoted by $0\leq n\leq 1$, such that
\begin{eqnarray}
	\label{eq:coevolution1}
	\epsilon\dot{x} &=&  x(1-x)\left[r_C(x,A(n))-r_D(x,A(n))\right], \label{eq.dxdt} \\
	\label{eq:coevolution2}
	\dot{n} &=&  n(1-n)\left(\theta x - (1-x)\right). \label{eq.dndt}
\end{eqnarray}
where $\epsilon$ is a speed parameter and $\theta$ denotes the strength of
cooperators in restoring the environment. 
In this coevolutionary model, the payoff matrices
$A_0$ and $A_1$ can have markedly different Nash equilibria \cite{nash1950bargaining}.
For example, when defection is uniformly favored when $n=1$
and cooperation is favored when $n=0$, then the 
the system can
exhibit a novel phenomenon termed an `oscillatory
tragedy of the commons' (o-TOC). An o-TOC denotes
a trajectory in the phase plan that approaches a heteroclinic cycle.
Given a replete environment, the population
rapidly switches from cooperation to defection, which then
degrades the environment. In the depleted environment,
cooperators re-establish, improving the environment,
then defectors invade and the cycle repeats.
Other outcomes, including a TOC and the aversion of a TOC
can emerge given other payoff matrices \cite{WeitzE7518}.

This coevolutionary game model is the basis for our development
and analysis of an individual-based framework to assess the influence of noise (first) and spatially explicit interactions (second) on the emergent dynamics of social context and resources.
To begin, consider a system comprised of
$N_C$ cooperators and $N_D$ defectors, such that $N=N_C+N_D$. 
A single time step consists of $N$ events.  In each event, a randomly chosen 
individual (the focal player) interacts with another
individual (the opponent) chosen at random. The payoff to the focal
player influences its probability to reproduce. 
Critically in our proposed framework, successful 
reproduction by the focal player replaces a randomly chosen \emph{third} individual
(see~\cite{bauer2018multiple} for a related public goods model that decouples interaction
and reproduction).
The following reactions denote those transitions that lead to a change in the
number of cooperators or defectors:
\begin{equation}\label{eq:reactions}
	\begin{aligned}
		&\overbrace{C}^{\text{focal}} &+
		&\overbrace{C}^{\text{opponent}}\rule{-12pt}{0ex}&+
		&\overbrace{D}^{\substack{\text{random} \\ \text{player}}}\rule{-12pt}{0ex}
		&\xrightarrow{k_1}
		&\overbrace{C}^{\text{focal}} &+ 
		&\overbrace{C}^{\text{opponent}}\rule{-12pt}{0ex} &+
		&\overbrace{C}^{\text{replaced}} \\
		&\rule{7pt}{0ex}C &+ &\rule{13pt}{0ex}D &+ &\rule{12pt}{0ex}D &\xrightarrow{k_2} &\rule{7pt}{0ex}C &+ &\rule{13pt}{0ex}D &+ &\rule{12pt}{0ex}C \\
		&\rule{7pt}{0ex}D &+ &\rule{13pt}{0ex}C &+ &\rule{12pt}{0ex}C &\xrightarrow{k_3} &\rule{7pt}{0ex}D &+ &\rule{13pt}{0ex}C &+ &\rule{12pt}{0ex}D \\
		&\rule{7pt}{0ex}D &+ &\rule{13pt}{0ex}D &+ &\rule{12pt}{0ex}C &\xrightarrow{k_4} &\rule{7pt}{0ex}D &+ &\rule{13pt}{0ex}D &+ &\rule{12pt}{0ex}D \\
	\end{aligned}
\end{equation}
where $k_i$ denote reproduction rates.

In the three-individual framework,
the master equation for the dynamics of cooperators is:
\begin{equation}
\begin{aligned}
\label{eq:main_master}
\mathbb{P}\left(n_C, \tau +\Delta \tau \right)&= \mathbb{P}\left(n_C, \tau \right) 
  + \mathbb{T}(n_C | n_C - 1 )\mathbb{P}(n_C- 1, \tau) \Delta \tau  \\
&+ \mathbb{T}(n_C | n_C + 1 )\mathbb{P}(n_C + 1, \tau) \Delta \tau \\
& - \mathbb{T}(n_C+1 | n_C)\mathbb{P}(n_C, \tau) \Delta \tau \\
&- \mathbb{T}(n_C-1 | n_C)\mathbb{P}(n_C, \tau)\Delta \tau + \mathcal{O}(\Delta \tau^2), 
\end{aligned}
\end{equation}
where the transition rates are:
\begin{equation}\label{eq:trans_rates}
\begin{split}
\mathbb{T}(n_C+1 | n_C) &= k_1 n_C   
\frac{n_C-1}{N-1} \frac{n_D}{N-1}  
+ k_2   n_C  \frac{n_D}{N-1}  \frac{n_D}{N-1} \\
		\\
\mathbb{T}(n_C-1 | n_C) &= k_3   n_D  \frac{n_C}{N-1} \frac{n_C}{N-1} 
+ k_4   n_D    \frac{n_D-1}{N-1} \frac{n_C}{N-1}.
\end{split}
\end{equation}
In the SI, we derive
the expected mean field dynamics 
for the frequency of cooperators
$x\equiv \lim_{N,n_c\rightarrow\infty} \left(\frac{n_c}{N}\right)$
from the master equation:
\begin{equation}
\begin{aligned}\label{eq:avg_dot2}
	\dot{x} &= x(1-x) \left[ (k_1-k_3)x + (k_2-k_4)(1-x)\right].
\end{aligned}
\end{equation}
We recover the replicator dynamics of the coevolutionary model
when
$k_1=R(n)$,
$k_2=S(n)$, $k_3=T(n)$, and $k_4=P(n)$. Hence,
transition rates are a function of
resource- and social-context dependent payoffs. 
In contrast, mean field dynamics derived via a two-player individual based model formulation (IBM2) 
result in a logistic dependency on $x$ distinct from the cubic nonlinearity
in Eq.~\ref{eq:avg_dot2} (see SI for derivation and details).

In order to further evaluate stochastic
dynamics of the IBM formulation, we simulated the joint dynamics of resources $n$ and social
context $x$ using $N=10^4$ individuals. A single time step over
an interval $\Delta t$
includes $N$ game steps followed by changes in
resource levels, $n(t)$ according to Eq.~\ref{eq.dndt} (see Supplementary
Information (SI) for details).  Given the master equation analysis,
we define reproduction rates $k_i$ based on the current environmental
state $n(t)$.
Consistent with our finding from the master equation,
the simulation results of the individual-based model involving three players (IBM3) 
recapitulate predictions of the mean-field replicator dynamics
model  (see FIG.~\ref{fig:IBM_ODE}-right). 
Specifically, we identify seven distinct phases corresponding to the relative
magnitude of payoffs given the resource deplete state. The
phases and their asymptotic behavior agree qualitatively 
with mean-field predictions.  In contrast,
if the focal player reproduces and replaces the opponent (as is
often assumed in two-player variants of spatial games), then 
the individual-based simulations diverge from predictions 
(see FIG.~\ref{fig:IBM_ODE}-left) as anticipated from expected
mean field dynamics (see SI).

There are two notable quantitative differences in the IBM3
simulations with respect to predictions from replicator dynamics.
First, whereas mean-field dynamics predict convergence to a heteroclinic
cycle (see `o-TOC' region in FIG.~\ref{fig:IBM_ODE}-right), the 
IBM simulations stochastically reach an absorbing state on the boundary.  
Such a result
is anticipated in any finite size simulation, given that heteroclinic cycles
asymptotically approach the boundary.  Second, the mean field
model predicts closed period orbits given certain
symmetric properties of $A_0$ and $A_1$ (corresponding to the
line with slope $(T_1-R_1)/(P_1-S_1)$ in FIG.~\ref{fig:IBM_ODE}-right.)
In contrast, the IBM simulations have demographic noise, which can
lead to repeated oscillations and convergence to a 
boundary (see FIG. S3).

To study the combined effects of spatial structure and demographic noise (see~\cite{durrett1994importance})
we extended the IBM3 framework
a 2-dimensional fully occupied lattice with $L$ sites per
dimension given periodic boundary conditions, where
the $N=L^2$ individuals  are either 
cooperators or defectors.
The focal player is selected at random
and opponent is chosen
randomly from the von Neumann neighborhood of the focal player.
We denote the position of the focal player (opponent) 
as $\vec{r}_F$ ($\vec{r}_O$).
The focal player reproduces with probability rate $k_m(s_F, s_O, \bar{n})$ given
the strategy set of focal player and opponent, $s_F$ and $s_O$, and the average
local environment, \( \bar{n} = (n(\vec{r}_F) + n(\vec{r}_O))/2\).  
Environmental state dynamics $n(\vec{r}, t)$ are augmented
by diffusion, i.e.,:
\begin{equation}
	\frac{\partial n}{\partial t} =  n(1-n)\left(\theta x - (1-x)\right) +D_n\nabla^2 n \label{eq.dndt_space}.
\end{equation}
The diffusivity $D_n$ controls the 
redistribution of resources relative to population dynamics.

\begin{figure}[t!]
\begin{center}
	\includegraphics[width=0.225\textwidth]{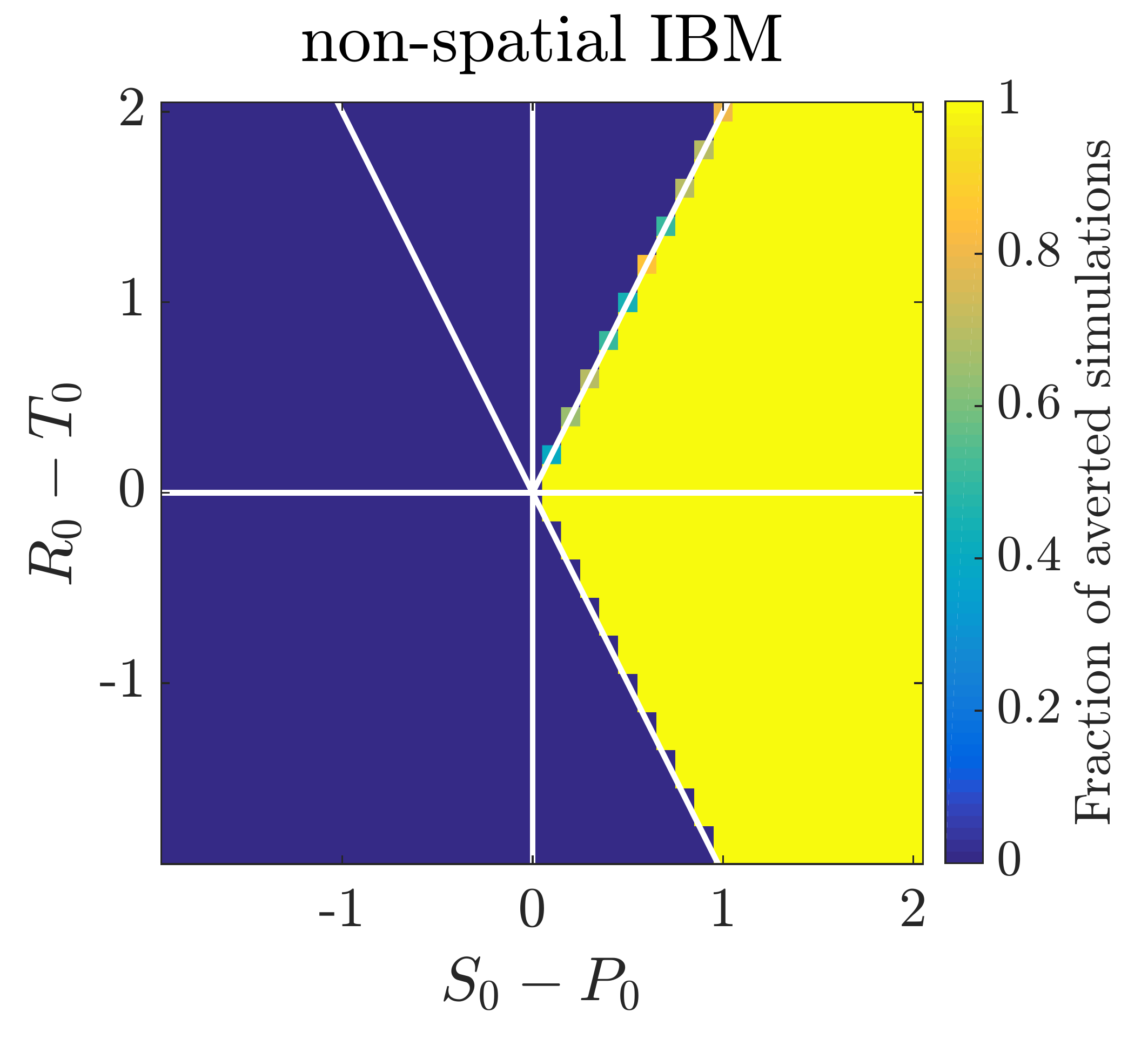}\\
	\includegraphics[width=0.15\textwidth]{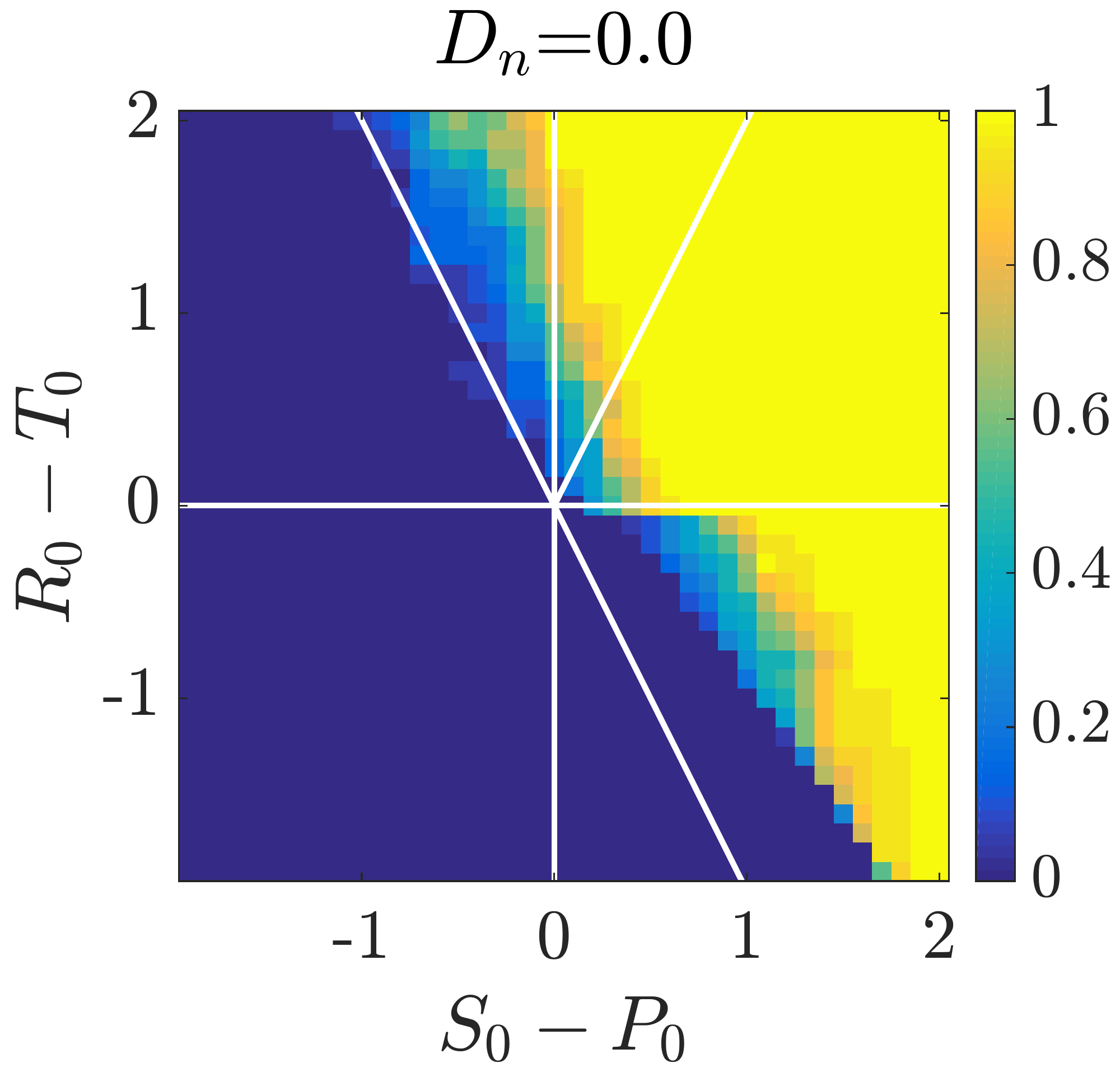}
	\includegraphics[width=0.15\textwidth]{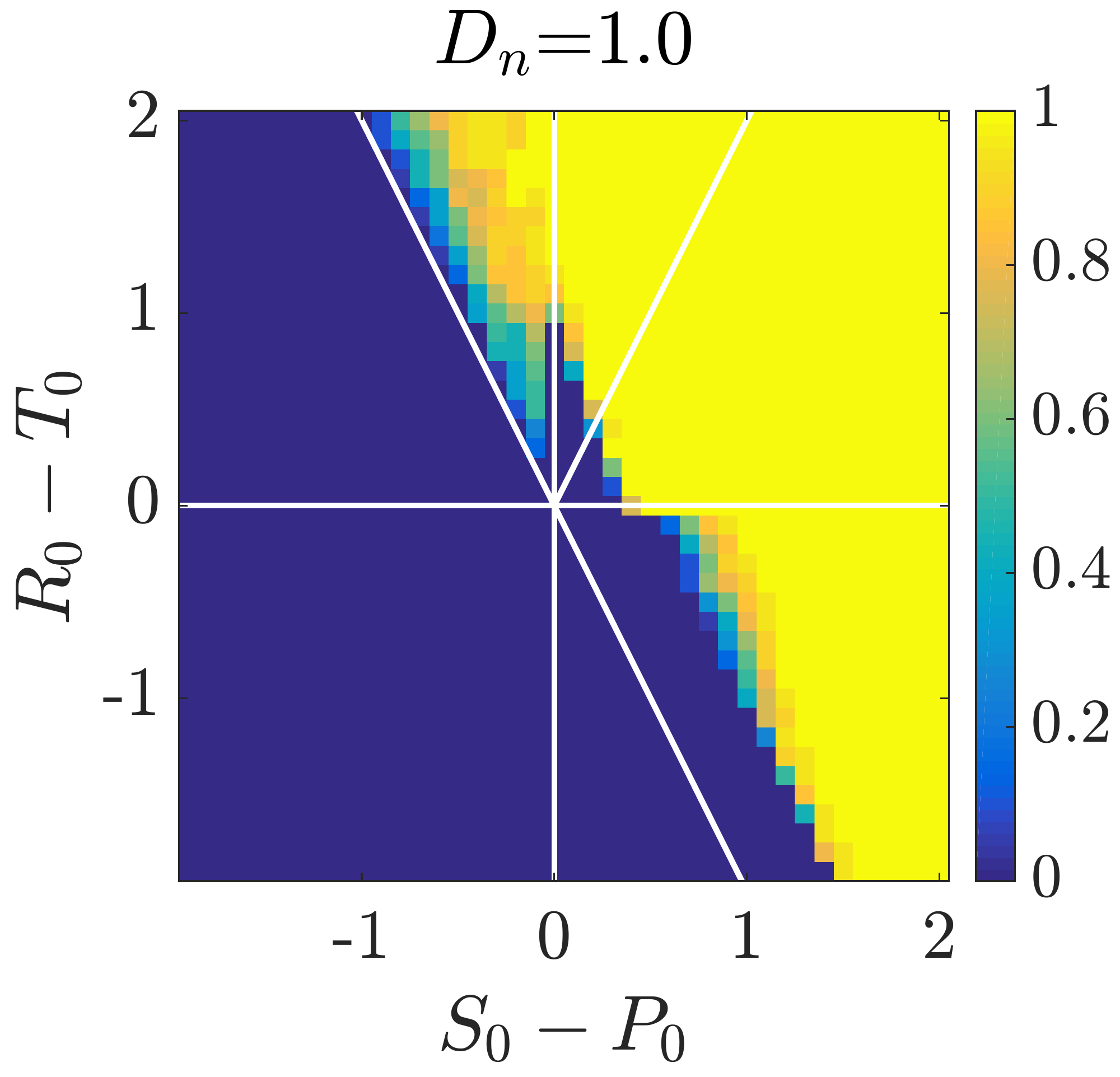}
	\includegraphics[width=0.15\textwidth]{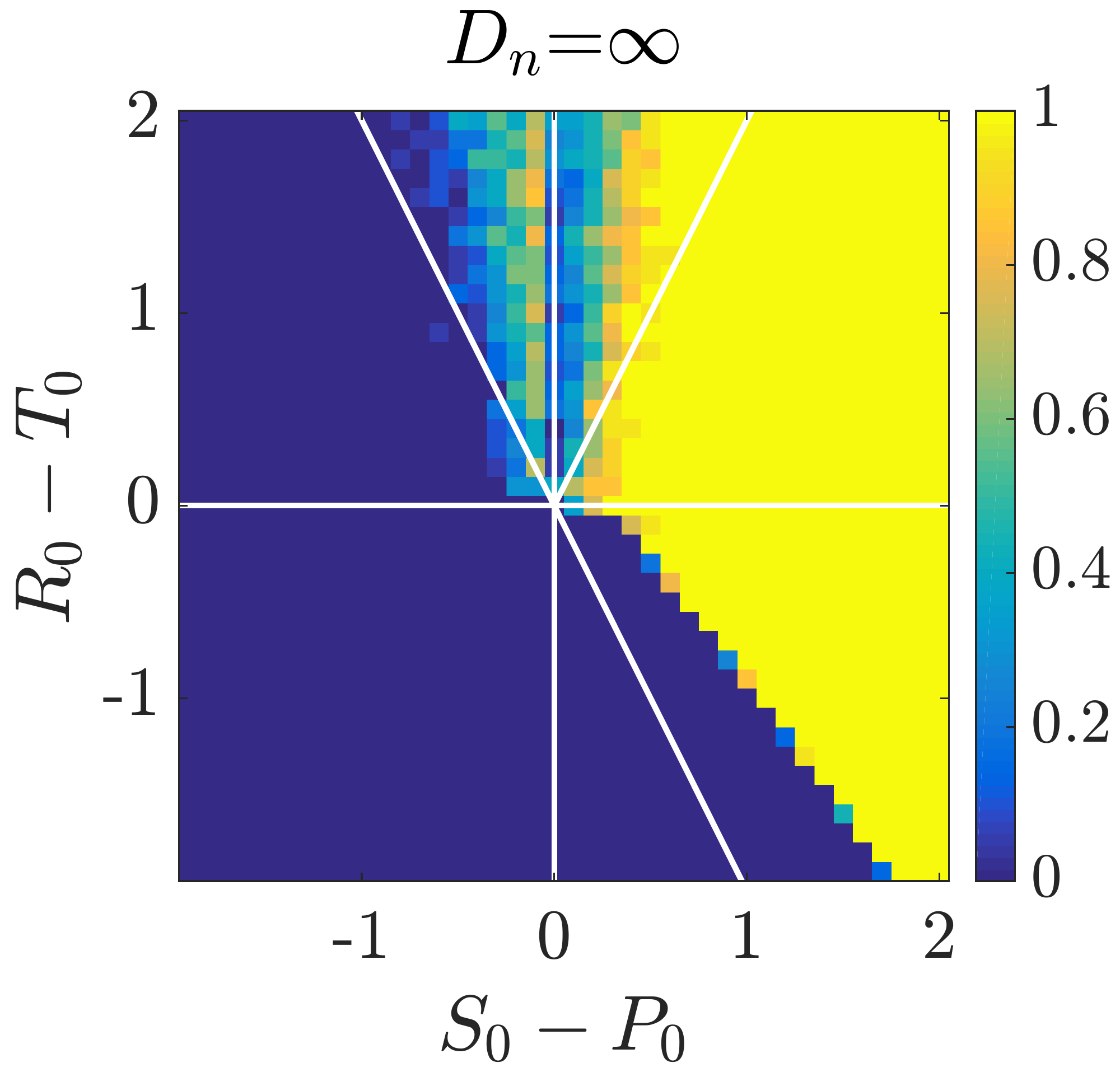}
		\caption{\footnotesize Strategy-resource dynamics given spatial interactions.  Colors in each heat-map denote the fraction of averted dynamics out of 20 replicates with different $A_0$'s.  The horizontal axis of the heat maps are $S_0 - P_0$ and the vertical ones are $R_0 - T_0$.  Each grid on the heat maps has increment $0.1$. The diffusivity $D_n$ is showed in the title of each panel. Other parameters for all replicates are $L=100$, $\theta=2$, $\epsilon = 0.5$, $\Delta x =1$, and $\Delta t = 0.05$, $A_1 = [3, 0;5, 1]$.  The white lines mark out the boundary of different dynamics predicted by the mean field model. Full parameter list for $A_0$ in each regime are listed in FIG. S2.}		
		\label{fig2.spatial_phase}
\end{center}
\vspace{-0.2in}
\end{figure}

Simulations of coevolutionary game-environmental dynamics
reveal dramatic changes in outcomes given spatially explicit interactions.
FIG.~\ref{fig2.spatial_phase} compares
dynamics of non-spatial and spatial IBM models with three different
diffusivities, $D_n=0, 1, \infty$, classifying outcomes
based on whether there is a TOC or not (the latter 
we term averted, see SI for criteria).
The heat maps show the proportion of averted cases among all replicates.
Spatial interactions enable TOC aversion when cooperation is favored 
given a coordination game context ($R_0>T_0$ and $S_0<P_0$, see upper left).  
However, spatial interactions also restrict
the parameter regimes where a TOC can be averted given
an anti-coordination game context ($R_0<T_0$ and
$S_0>P_0$, see bottom right).
For long-term dynamics, we find that
oscillating dynamics are typical in
$D_n=\infty$ cases (see examples in the SI).
Such oscillatory dynamics can spiral inwards when
TOC-s are averted or outwards to the boundary. Of note,
amongst IBM models we only observe a persistent o-TOC when
$D_n=\infty$; indicating the role of strong spatial coupling 
to induce oscillations.

We further investigated spatiotemporal dynamics focusing
on variation
in $D_n$ given parameter regimes with both averted and TOC dynamics.
These regimes correspond to the case where $S_0<P_0$, $R_0>T_0$ and
where $R_0>-\theta (S_0 - P_0)+T_0$ (see bottom panels of FIG.~\ref{fig2.spatial_phase}).
The results of spatially explicit IBM3 model simulations 
are shown in FIG. 3
for $D_n=0, 1$ and $\infty$.
Notably, all cases appear to exhibit clustering amongst
cooperators and the cases with heterogeneous
environmental dynamcis ($D_n=0$ and $D_n=1$) also
appear to exhibit clustering between
cooperators and environmental resource state.
However, there are markedly different types of emergent spatial
patterns give variation in the diffusivity of environmental resource
state. In order to assess clustering quantitatively, we analyzed 
the joint structure
of social context and resource levels by measuring the
spatial cross-correlation function:
\begin{equation}
	g_{CN}(r,t) = \frac{L^2}{\mathcal{A}(r)}\frac{\Sigma_{i, j}(\Sigma_{i', j'} x_{i,j}(t)\cdot n_{i',j'}(t))}{\Sigma_{i, j} x_{i,j}(t) \Sigma_{i, j} n_{i,j}(t)}, \\
\end{equation}
and the spatial autocorrelation function of cooperator clustering:
\begin{equation}
	g_{CC}(r,t) =  \frac{L^2}{\mathcal{A}(r)}\frac{\Sigma_{i, j}(\Sigma_{i', j'} x_{i,j}(t)\cdot x_{i',j'}(t))}{(\Sigma_{i, j} x_{i,j}(t))^2}, \\
\end{equation}
where $ r < \sqrt{(i-i')^2+(j-j')^2} \leq r+1$, and $\mathcal{A}(r)$ denotes the
 number of lattice sites within this range in both cases.
 We then fit the short-range components
of the observed correlation at a fixed
time point to a decaying exponential,
i.e., $g(r,t)\sim 1+\alpha(t)e^{-r/\xi(t)}$ given
pre-factor $\alpha$ and correlation length $\xi$.We then fit the short-range components
of the observed correlation at a fixed
time point to a decaying exponential,
i.e., $g(r,t)\sim 1+\alpha(t)e^{-r/\xi(t)}$ given
pre-factor $\alpha$ and correlation length $\xi$.
The spatial autocorrelation analysis confirms the
emergence of clustering amongst cooperators
when the TOC is averted, i.e., $g_{CC}(r)>1$ for $r\rightarrow 1$
(see black lines in the sub-panels of FIG. 3).
Yet there are marked differences in the dynamics of the
cross-correlation between cooperators and the environmental state.

For $D_n=0$, the environment and cooperative population propagate outward as a wave. 
The cooperative population spread leaving patches of
resource replete environments.
The $g_{CN}(r)$ plots shows that
$x$ and $n$ can be positively correlated as a wave initiates but negatively 
correlated once defectors invade and replace resource replete
environments, leading to (often disjoint) patchy distributions of both
resources and cooperators. 
In contrast, for $D_n=1$, small clusters of cooperators
and localized resources form after
initial transient dynamics. 
This feature is captured by the $g_{CN}(r)$ analysis, revealing
strongly elevated cross-correlation (see the middle row of FIG. 3) as well as similar pattern in the dynamics of
$g_{CC}(r)$ and $g_{CN}(r)$. 
We note that these `gangs' of cooperators and their environmental `tail' are chased by a dominant group of 
defectors (see~\citep{szabo_pre1998} for related findings in
evolutionary PD models withoout environmental feedback).
Finally, given $D_n=\infty$, the resources are uniform across space.
Cooperative clusters grow towards system sizes due to the strong spatial coupling mediated via fast resource
diffusivity. The single large cooperator cluster expands and shrinks over time with increasing amplitude, 
as evidenced by the elevated autocorrelation of $g_{CC}(r)$ in the bottom row of FIG. 3, with rapid switches in resource state, leading to an eventual collapse
of the cooperator population.  We do not report $g_{CN}(r)$ 
given the uniform distribution of resources given $D_n=\infty$.

\begin{figure}[t!]
	\begin{center}
		\includegraphics[width=0.48\textwidth]{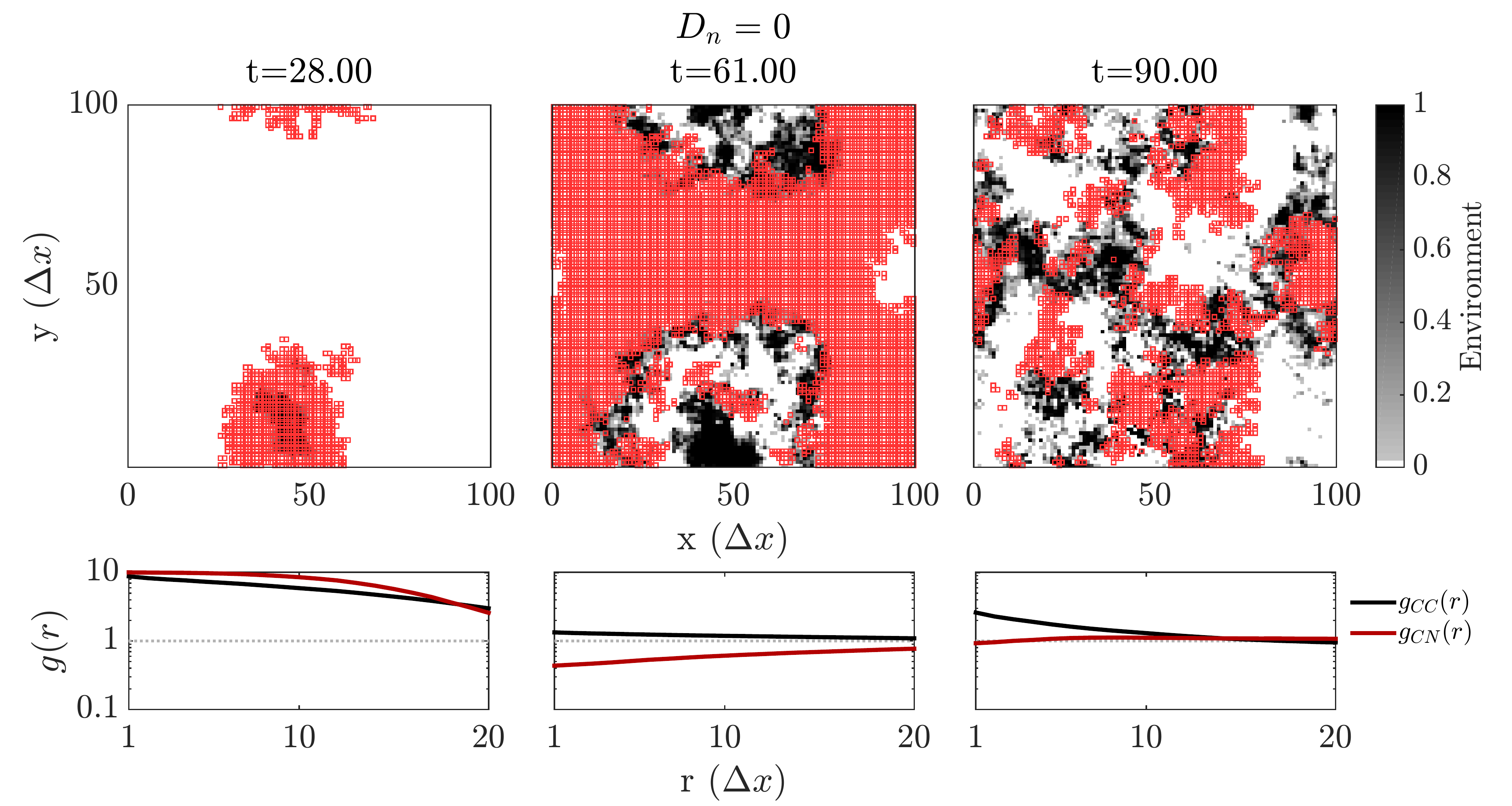}
		\includegraphics[width=0.48\textwidth]{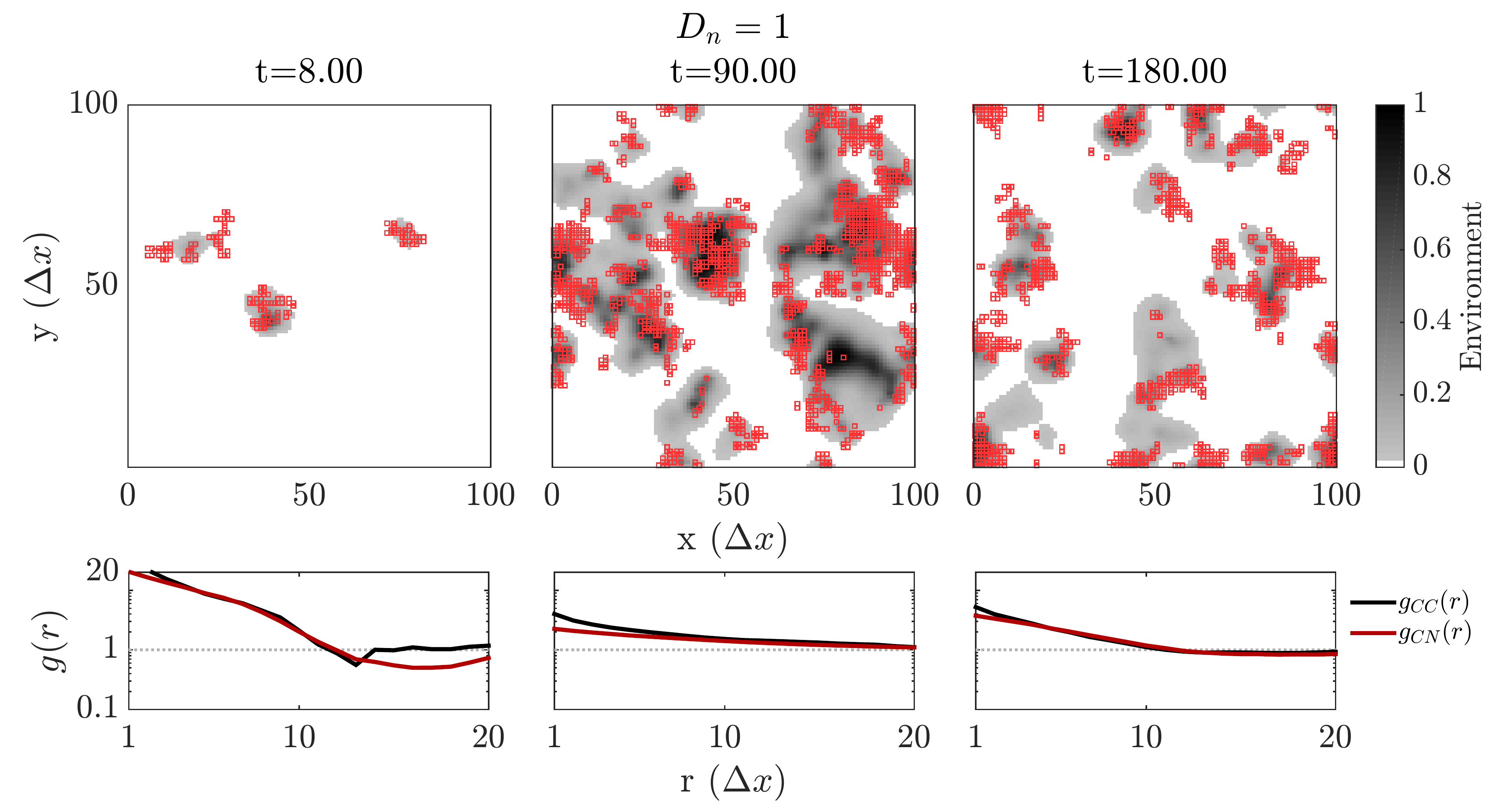}
		\includegraphics[width=0.48\textwidth]{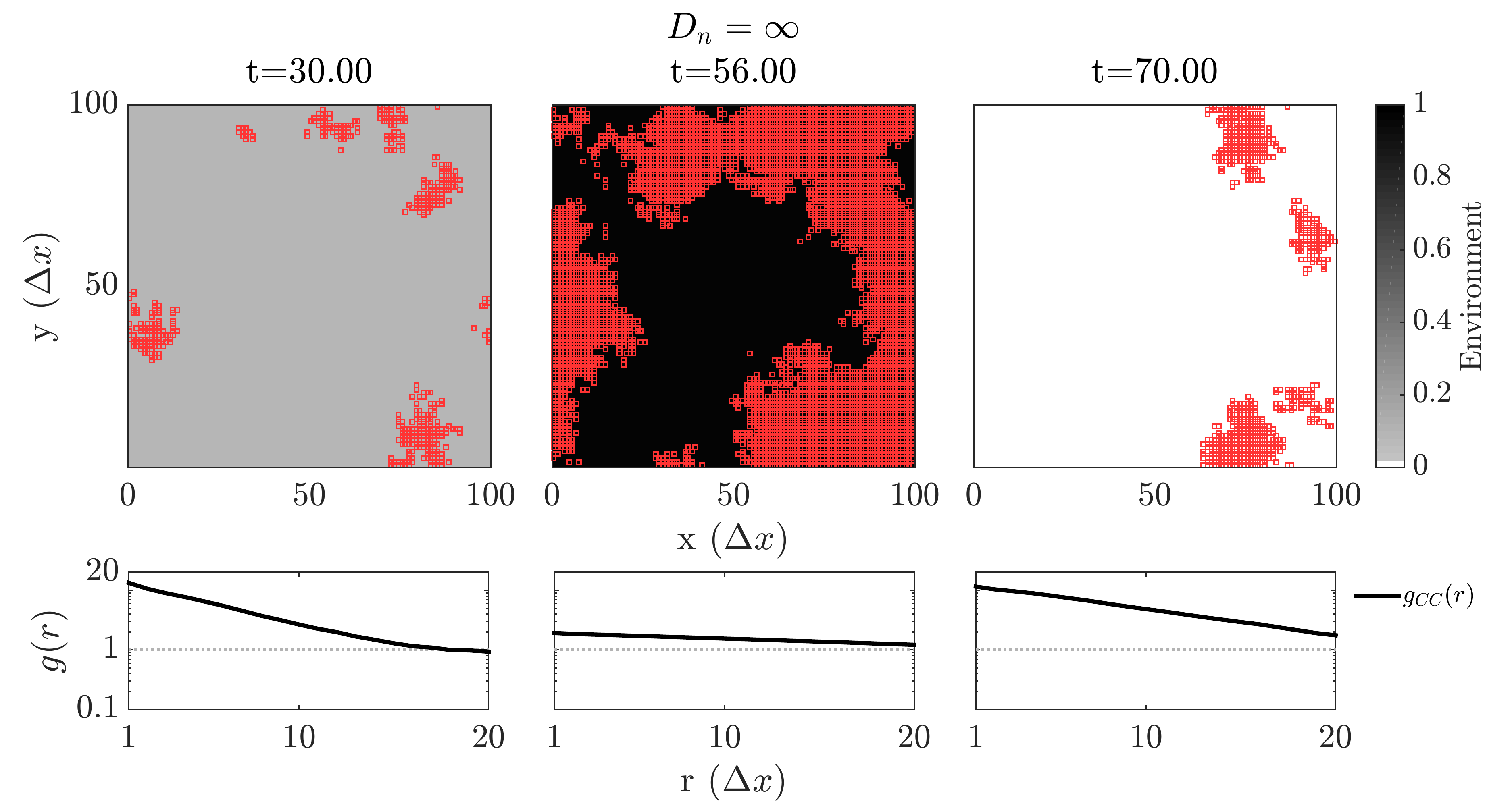}
		\caption{\footnotesize Spatiotemporal dynamics of resources and cooperation.
		The background color represents the environment,
		while a red square means a cooperator occupy the lattice site. The empty sites are occupied by defectors.
		(Top row) $D_n=0$, a circular wave of cooperative population propagates outward.
		 (Middle row) $D_n=1$, a few small patches of cooperators move around and divide.
		  (Bottom row) $D_n=\infty$, a large cooperator cluster expand and shrink over time with increasing amplitude until extinction.}
	  \label{fig3.pair_correlation}
	\end{center}
\end{figure}

In summary, we have developed an individual-based framework
to incorporate the analysis of demographic noise and spatial
interactions in coevolutionary game dynamics that coupled
individual strategies and the environment \cite{durrett1994importance}.
The IBM involving 3 players in a game recapitulates and generalizes
earlier findings from a coevolutionary game model, including
the emergence of an oscillatory tragedy of the commons.
Spatial interactions can shift the domains in which
a tragedy of the commons may arise when compared to non-spatial
models \cite{halatek2018rethinking}.  Spatially explicit dynamics
also lead to novel, coherent spatiotemporal patterns \cite{nowak1992evolutionary,butler2009robust,butler2011fluctuation,nowak1993spatial,nanda2017spatial}, 
including diffusive clusters, flickering, and wave-like patterns.
These joint dynamics of resources and social strategies suggest
multiple avenues for future study, including formally deriving
effective PDEs to characterize whether the system permits propagating waves
in the large system limit. It will also be critical
to evaluate the extent to which
spatial interactions modify strategy-environment feedback
in proposed generalizations~\cite{tilman_biorxiv} of the replicator
framework underlying the present work~\cite{WeitzE7518}
and in stochastic games with feedback between behavior and public good states~\citep{hilbe_nature2018}.
Finally, the spatial approach
may also aid efforts to understand
how microorganisms produce and utilize public goods, \emph{e.g.,}
siderophores -- extracellular iron harvesting enzymes \cite{cordero_pnas2012,niehus2017_evolution,bauer2018multiple,menon2015public,lewin2017microbes}. 
Given increasing pressures on limited resources, we intend to leverage
prior work on controlling mean-field strategy-environment dynamics \cite{paarporn2018optimal}
to identify ways in which local manipulation of resources, strategies,
and/or perceptions can help stabilize and conserve the commons.

\footnotesize
\section*{Acknowledgements}
We thank Erol Ak\c{c}ay, Marianne Bauer, Ceyhun Eksin, Guanlin Li, Keith Paarporn, William Ratcliff, and Peter Yunker for feedback and Stephen Beckett for reviewing code.
This work was supported by the Simons Foundation (SCOPE award ID 329108, J.S.W.) and by Army Research Office grant W911NF-14-1-0402.

\bibliography{refs}

\begin{thebibliography}{21}%
\makeatletter
\providecommand \@ifxundefined [1]{%
 \@ifx{#1\undefined}
}%
\providecommand \@ifnum [1]{%
 \ifnum #1\expandafter \@firstoftwo
 \else \expandafter \@secondoftwo
 \fi
}%
\providecommand \@ifx [1]{%
 \ifx #1\expandafter \@firstoftwo
 \else \expandafter \@secondoftwo
 \fi
}%
\providecommand \natexlab [1]{#1}%
\providecommand \enquote  [1]{``#1''}%
\providecommand \bibnamefont  [1]{#1}%
\providecommand \bibfnamefont [1]{#1}%
\providecommand \citenamefont [1]{#1}%
\providecommand \href@noop [0]{\@secondoftwo}%
\providecommand \href [0]{\begingroup \@sanitize@url \@href}%
\providecommand \@href[1]{\@@startlink{#1}\@@href}%
\providecommand \@@href[1]{\endgroup#1\@@endlink}%
\providecommand \@sanitize@url [0]{\catcode `\\12\catcode `\$12\catcode
  `\&12\catcode `\#12\catcode `\^12\catcode `\_12\catcode `\%12\relax}%
\providecommand \@@startlink[1]{}%
\providecommand \@@endlink[0]{}%
\providecommand \url  [0]{\begingroup\@sanitize@url \@url }%
\providecommand \@url [1]{\endgroup\@href {#1}{\urlprefix }}%
\providecommand \urlprefix  [0]{URL }%
\providecommand \Eprint [0]{\href }%
\providecommand \doibase [0]{http://dx.doi.org/}%
\providecommand \selectlanguage [0]{\@gobble}%
\providecommand \bibinfo  [0]{\@secondoftwo}%
\providecommand \bibfield  [0]{\@secondoftwo}%
\providecommand \translation [1]{[#1]}%
\providecommand \BibitemOpen [0]{}%
\providecommand \bibitemStop [0]{}%
\providecommand \bibitemNoStop [0]{.\EOS\space}%
\providecommand \EOS [0]{\spacefactor3000\relax}%
\providecommand \BibitemShut  [1]{\csname bibitem#1\endcsname}%
\let\auto@bib@innerbib\@empty
\bibitem [{\citenamefont {Hardin}(1968)}]{Hardin1243}%
  \BibitemOpen
  \bibfield  {author} {\bibinfo {author} {\bibfnamefont {G.}~\bibnamefont
  {Hardin}},\ }\href {\doibase 10.1126/science.162.3859.1243} {\bibfield
  {journal} {\bibinfo  {journal} {Science}\ }\textbf {\bibinfo {volume}
  {162}},\ \bibinfo {pages} {1243} (\bibinfo {year} {1968})}\BibitemShut
  {NoStop}%
\bibitem [{\citenamefont {Nowak}(2006)}]{nowak2006evolutionary}%
  \BibitemOpen
  \bibfield  {author} {\bibinfo {author} {\bibfnamefont {M.~A.}\ \bibnamefont
  {Nowak}},\ }\href@noop {} {\emph {\bibinfo {title} {Evolutionary Dynamics}}}\
  (\bibinfo  {publisher} {Harvard University Press},\ \bibinfo {year}
  {2006})\BibitemShut {NoStop}%
\bibitem [{\citenamefont {Smith}\ and\ \citenamefont
  {Price}(1973)}]{smith1973logic}%
  \BibitemOpen
  \bibfield  {author} {\bibinfo {author} {\bibfnamefont {J.~M.}\ \bibnamefont
  {Smith}}\ and\ \bibinfo {author} {\bibfnamefont {G.~R.}\ \bibnamefont
  {Price}},\ }\href@noop {} {\bibfield  {journal} {\bibinfo  {journal}
  {Nature}\ }\textbf {\bibinfo {volume} {246}},\ \bibinfo {pages} {15}
  (\bibinfo {year} {1973})}\BibitemShut {NoStop}%
\bibitem [{\citenamefont {Weitz}\ \emph {et~al.}(2016)\citenamefont {Weitz},
  \citenamefont {Eksin}, \citenamefont {Paarporn}, \citenamefont {Brown},\ and\
  \citenamefont {Ratcliff}}]{WeitzE7518}%
  \BibitemOpen
  \bibfield  {author} {\bibinfo {author} {\bibfnamefont {J.~S.}\ \bibnamefont
  {Weitz}}, \bibinfo {author} {\bibfnamefont {C.}~\bibnamefont {Eksin}},
  \bibinfo {author} {\bibfnamefont {K.}~\bibnamefont {Paarporn}}, \bibinfo
  {author} {\bibfnamefont {S.~P.}\ \bibnamefont {Brown}}, \ and\ \bibinfo
  {author} {\bibfnamefont {W.~C.}\ \bibnamefont {Ratcliff}},\ }\href {\doibase
  10.1073/pnas.1604096113} {\bibfield  {journal} {\bibinfo  {journal}
  {Proceedings of the National Academy of Sciences}\ }\textbf {\bibinfo
  {volume} {113}},\ \bibinfo {pages} {E7518} (\bibinfo {year}
  {2016})}\BibitemShut {NoStop}%
\bibitem [{\citenamefont {Nash~Jr}(1950)}]{nash1950bargaining}%
  \BibitemOpen
  \bibfield  {author} {\bibinfo {author} {\bibfnamefont {J.~F.}\ \bibnamefont
  {Nash~Jr}},\ }\href@noop {} {\bibfield  {journal} {\bibinfo  {journal}
  {Econometrica: Journal of the Econometric Society}\ ,\ \bibinfo {pages}
  {155}} (\bibinfo {year} {1950})}\BibitemShut {NoStop}%
\bibitem [{\citenamefont {Bauer}\ and\ \citenamefont
  {Frey}(2018)}]{bauer2018multiple}%
  \BibitemOpen
  \bibfield  {author} {\bibinfo {author} {\bibfnamefont {M.}~\bibnamefont
  {Bauer}}\ and\ \bibinfo {author} {\bibfnamefont {E.}~\bibnamefont {Frey}},\
  }\href@noop {} {\bibfield  {journal} {\bibinfo  {journal} {Physical Review
  E}\ }\textbf {\bibinfo {volume} {97}},\ \bibinfo {pages} {042307} (\bibinfo
  {year} {2018})}\BibitemShut {NoStop}%
\bibitem [{\citenamefont {Durrett}\ and\ \citenamefont
  {Levin}(1994)}]{durrett1994importance}%
  \BibitemOpen
  \bibfield  {author} {\bibinfo {author} {\bibfnamefont {R.}~\bibnamefont
  {Durrett}}\ and\ \bibinfo {author} {\bibfnamefont {S.}~\bibnamefont
  {Levin}},\ }\href@noop {} {\bibfield  {journal} {\bibinfo  {journal}
  {Theoretical Population Biology}\ }\textbf {\bibinfo {volume} {46}},\
  \bibinfo {pages} {363} (\bibinfo {year} {1994})}\BibitemShut {NoStop}%
\bibitem [{\citenamefont {Szab{\'o}}\ and\ \citenamefont
  {T{\H{o}}ke}(1998)}]{szabo_pre1998}%
  \BibitemOpen
  \bibfield  {author} {\bibinfo {author} {\bibfnamefont {G.}~\bibnamefont
  {Szab{\'o}}}\ and\ \bibinfo {author} {\bibfnamefont {C.}~\bibnamefont
  {T{\H{o}}ke}},\ }\href@noop {} {\bibfield  {journal} {\bibinfo  {journal}
  {Physical Review E}\ }\textbf {\bibinfo {volume} {58}},\ \bibinfo {pages}
  {69} (\bibinfo {year} {1998})}\BibitemShut {NoStop}%
\bibitem [{\citenamefont {Halatek}\ and\ \citenamefont
  {Frey}(2018)}]{halatek2018rethinking}%
  \BibitemOpen
  \bibfield  {author} {\bibinfo {author} {\bibfnamefont {J.}~\bibnamefont
  {Halatek}}\ and\ \bibinfo {author} {\bibfnamefont {E.}~\bibnamefont {Frey}},\
  }\href@noop {} {\bibfield  {journal} {\bibinfo  {journal} {Nature Physics}\
  }\textbf {\bibinfo {volume} {14}},\ \bibinfo {pages} {507} (\bibinfo {year}
  {2018})}\BibitemShut {NoStop}%
\bibitem [{\citenamefont {Nowak}\ and\ \citenamefont
  {May}(1992)}]{nowak1992evolutionary}%
  \BibitemOpen
  \bibfield  {author} {\bibinfo {author} {\bibfnamefont {M.~A.}\ \bibnamefont
  {Nowak}}\ and\ \bibinfo {author} {\bibfnamefont {R.~M.}\ \bibnamefont
  {May}},\ }\href@noop {} {\bibfield  {journal} {\bibinfo  {journal} {Nature}\
  }\textbf {\bibinfo {volume} {359}},\ \bibinfo {pages} {826} (\bibinfo {year}
  {1992})}\BibitemShut {NoStop}%
\bibitem [{\citenamefont {Butler}\ and\ \citenamefont
  {Goldenfeld}(2009)}]{butler2009robust}%
  \BibitemOpen
  \bibfield  {author} {\bibinfo {author} {\bibfnamefont {T.}~\bibnamefont
  {Butler}}\ and\ \bibinfo {author} {\bibfnamefont {N.}~\bibnamefont
  {Goldenfeld}},\ }\href@noop {} {\bibfield  {journal} {\bibinfo  {journal}
  {Physical Review E}\ }\textbf {\bibinfo {volume} {80}},\ \bibinfo {pages}
  {030902} (\bibinfo {year} {2009})}\BibitemShut {NoStop}%
\bibitem [{\citenamefont {Butler}\ and\ \citenamefont
  {Goldenfeld}(2011)}]{butler2011fluctuation}%
  \BibitemOpen
  \bibfield  {author} {\bibinfo {author} {\bibfnamefont {T.}~\bibnamefont
  {Butler}}\ and\ \bibinfo {author} {\bibfnamefont {N.}~\bibnamefont
  {Goldenfeld}},\ }\href@noop {} {\bibfield  {journal} {\bibinfo  {journal}
  {Physical Review E}\ }\textbf {\bibinfo {volume} {84}},\ \bibinfo {pages}
  {011112} (\bibinfo {year} {2011})}\BibitemShut {NoStop}%
\bibitem [{\citenamefont {Nowak}\ and\ \citenamefont
  {May}(1993)}]{nowak1993spatial}%
  \BibitemOpen
  \bibfield  {author} {\bibinfo {author} {\bibfnamefont {M.~A.}\ \bibnamefont
  {Nowak}}\ and\ \bibinfo {author} {\bibfnamefont {R.~M.}\ \bibnamefont
  {May}},\ }\href@noop {} {\bibfield  {journal} {\bibinfo  {journal}
  {International Journal of Bifurcation and Chaos}\ }\textbf {\bibinfo {volume}
  {3}},\ \bibinfo {pages} {35} (\bibinfo {year} {1993})}\BibitemShut {NoStop}%
\bibitem [{\citenamefont {Nanda}\ and\ \citenamefont
  {Durrett}(2017)}]{nanda2017spatial}%
  \BibitemOpen
  \bibfield  {author} {\bibinfo {author} {\bibfnamefont {M.}~\bibnamefont
  {Nanda}}\ and\ \bibinfo {author} {\bibfnamefont {R.}~\bibnamefont
  {Durrett}},\ }\href {\doibase 10.1073/pnas.1620852114} {\bibfield  {journal}
  {\bibinfo  {journal} {Proceedings of the National Academy of Sciences}\
  }\textbf {\bibinfo {volume} {114}},\ \bibinfo {pages} {6046} (\bibinfo {year}
  {2017})}\BibitemShut {NoStop}%
\bibitem [{\citenamefont {Tilman}\ \emph {et~al.}(2018)\citenamefont {Tilman},
  \citenamefont {Akcay},\ and\ \citenamefont {Plotkin}}]{tilman_biorxiv}%
  \BibitemOpen
  \bibfield  {author} {\bibinfo {author} {\bibfnamefont {A.~R.}\ \bibnamefont
  {Tilman}}, \bibinfo {author} {\bibfnamefont {E.}~\bibnamefont {Akcay}}, \
  and\ \bibinfo {author} {\bibfnamefont {J.}~\bibnamefont {Plotkin}},\ }\href
  {\doibase 10.1101/493023} {\bibfield  {journal} {\bibinfo  {journal}
  {bioRxiv}\ } (\bibinfo {year} {2018}),\ 10.1101/493023}\BibitemShut {NoStop}%
\bibitem [{\citenamefont {Hilbe}\ \emph {et~al.}(2018)\citenamefont {Hilbe},
  \citenamefont {{\v{S}}imsa}, \citenamefont {Chatterjee},\ and\ \citenamefont
  {Nowak}}]{hilbe_nature2018}%
  \BibitemOpen
  \bibfield  {author} {\bibinfo {author} {\bibfnamefont {C.}~\bibnamefont
  {Hilbe}}, \bibinfo {author} {\bibfnamefont {{\v{S}}.}~\bibnamefont
  {{\v{S}}imsa}}, \bibinfo {author} {\bibfnamefont {K.}~\bibnamefont
  {Chatterjee}}, \ and\ \bibinfo {author} {\bibfnamefont {M.~A.}\ \bibnamefont
  {Nowak}},\ }\href@noop {} {\bibfield  {journal} {\bibinfo  {journal}
  {Nature}\ }\textbf {\bibinfo {volume} {559}},\ \bibinfo {pages} {246}
  (\bibinfo {year} {2018})}\BibitemShut {NoStop}%
\bibitem [{\citenamefont {Cordero}\ \emph {et~al.}(2012)\citenamefont
  {Cordero}, \citenamefont {Ventouras}, \citenamefont {DeLong},\ and\
  \citenamefont {Polz}}]{cordero_pnas2012}%
  \BibitemOpen
  \bibfield  {author} {\bibinfo {author} {\bibfnamefont {O.~X.}\ \bibnamefont
  {Cordero}}, \bibinfo {author} {\bibfnamefont {L.-A.}\ \bibnamefont
  {Ventouras}}, \bibinfo {author} {\bibfnamefont {E.~F.}\ \bibnamefont
  {DeLong}}, \ and\ \bibinfo {author} {\bibfnamefont {M.~F.}\ \bibnamefont
  {Polz}},\ }\href@noop {} {\bibfield  {journal} {\bibinfo  {journal}
  {Proceedings of the National Academy of Sciences}\ }\textbf {\bibinfo
  {volume} {109}},\ \bibinfo {pages} {20059} (\bibinfo {year}
  {2012})}\BibitemShut {NoStop}%
\bibitem [{\citenamefont {Niehus}\ \emph {et~al.}(2017)\citenamefont {Niehus},
  \citenamefont {Picot}, \citenamefont {Oliveira}, \citenamefont {Mitri},\ and\
  \citenamefont {Foster}}]{niehus2017_evolution}%
  \BibitemOpen
  \bibfield  {author} {\bibinfo {author} {\bibfnamefont {R.}~\bibnamefont
  {Niehus}}, \bibinfo {author} {\bibfnamefont {A.}~\bibnamefont {Picot}},
  \bibinfo {author} {\bibfnamefont {N.~M.}\ \bibnamefont {Oliveira}}, \bibinfo
  {author} {\bibfnamefont {S.}~\bibnamefont {Mitri}}, \ and\ \bibinfo {author}
  {\bibfnamefont {K.~R.}\ \bibnamefont {Foster}},\ }\href@noop {} {\bibfield
  {journal} {\bibinfo  {journal} {Evolution}\ }\textbf {\bibinfo {volume}
  {71}},\ \bibinfo {pages} {1443} (\bibinfo {year} {2017})}\BibitemShut
  {NoStop}%
\bibitem [{\citenamefont {Menon}\ and\ \citenamefont
  {Korolev}(2015)}]{menon2015public}%
  \BibitemOpen
  \bibfield  {author} {\bibinfo {author} {\bibfnamefont {R.}~\bibnamefont
  {Menon}}\ and\ \bibinfo {author} {\bibfnamefont {K.~S.}\ \bibnamefont
  {Korolev}},\ }\href@noop {} {\bibfield  {journal} {\bibinfo  {journal}
  {Physical Review Letters}\ }\textbf {\bibinfo {volume} {114}},\ \bibinfo
  {pages} {168102} (\bibinfo {year} {2015})}\BibitemShut {NoStop}%
\bibitem [{\citenamefont {Lewin-Epstein}\ \emph {et~al.}(2017)\citenamefont
  {Lewin-Epstein}, \citenamefont {Aharonov},\ and\ \citenamefont
  {Hadany}}]{lewin2017microbes}%
  \BibitemOpen
  \bibfield  {author} {\bibinfo {author} {\bibfnamefont {O.}~\bibnamefont
  {Lewin-Epstein}}, \bibinfo {author} {\bibfnamefont {R.}~\bibnamefont
  {Aharonov}}, \ and\ \bibinfo {author} {\bibfnamefont {L.}~\bibnamefont
  {Hadany}},\ }\href@noop {} {\bibfield  {journal} {\bibinfo  {journal} {Nature
  Communications}\ }\textbf {\bibinfo {volume} {8}},\ \bibinfo {pages} {14040}
  (\bibinfo {year} {2017})}\BibitemShut {NoStop}%
\bibitem [{\citenamefont {Paarporn}\ \emph {et~al.}(2018)\citenamefont
  {Paarporn}, \citenamefont {Eksin}, \citenamefont {Weitz},\ and\ \citenamefont
  {Wardi}}]{paarporn2018optimal}%
  \BibitemOpen
  \bibfield  {author} {\bibinfo {author} {\bibfnamefont {K.}~\bibnamefont
  {Paarporn}}, \bibinfo {author} {\bibfnamefont {C.}~\bibnamefont {Eksin}},
  \bibinfo {author} {\bibfnamefont {J.~S.}\ \bibnamefont {Weitz}}, \ and\
  \bibinfo {author} {\bibfnamefont {Y.}~\bibnamefont {Wardi}},\ }\href@noop {}
  {\bibfield  {journal} {\bibinfo  {journal} {arXiv preprint arXiv:1803.06737}\
  } (\bibinfo {year} {2018})}\BibitemShut {NoStop}%
\end{thebibliography}%


\begin{thebibliography}{2}%
\makeatletter
\providecommand \@ifxundefined [1]{%
 \@ifx{#1\undefined}
}%
\providecommand \@ifnum [1]{%
 \ifnum #1\expandafter \@firstoftwo
 \else \expandafter \@secondoftwo
 \fi
}%
\providecommand \@ifx [1]{%
 \ifx #1\expandafter \@firstoftwo
 \else \expandafter \@secondoftwo
 \fi
}%
\providecommand \natexlab [1]{#1}%
\providecommand \enquote  [1]{``#1''}%
\providecommand \bibnamefont  [1]{#1}%
\providecommand \bibfnamefont [1]{#1}%
\providecommand \citenamefont [1]{#1}%
\providecommand \href@noop [0]{\@secondoftwo}%
\providecommand \href [0]{\begingroup \@sanitize@url \@href}%
\providecommand \@href[1]{\@@startlink{#1}\@@href}%
\providecommand \@@href[1]{\endgroup#1\@@endlink}%
\providecommand \@sanitize@url [0]{\catcode `\\12\catcode `\$12\catcode
  `\&12\catcode `\#12\catcode `\^12\catcode `\_12\catcode `\%12\relax}%
\providecommand \@@startlink[1]{}%
\providecommand \@@endlink[0]{}%
\providecommand \url  [0]{\begingroup\@sanitize@url \@url }%
\providecommand \@url [1]{\endgroup\@href {#1}{\urlprefix }}%
\providecommand \urlprefix  [0]{URL }%
\providecommand \Eprint [0]{\href }%
\providecommand \doibase [0]{http://dx.doi.org/}%
\providecommand \selectlanguage [0]{\@gobble}%
\providecommand \bibinfo  [0]{\@secondoftwo}%
\providecommand \bibfield  [0]{\@secondoftwo}%
\providecommand \translation [1]{[#1]}%
\providecommand \BibitemOpen [0]{}%
\providecommand \bibitemStop [0]{}%
\providecommand \bibitemNoStop [0]{.\EOS\space}%
\providecommand \EOS [0]{\spacefactor3000\relax}%
\providecommand \BibitemShut  [1]{\csname bibitem#1\endcsname}%
\let\auto@bib@innerbib\@empty
\bibitem [{\citenamefont {Press}\ \emph {et~al.}(2007)\citenamefont {Press},
  \citenamefont {Teukolsky}, \citenamefont {Vetterling},\ and\ \citenamefont
  {Flannery}}]{press2007numerical}%
  \BibitemOpen
  \bibfield  {author} {\bibinfo {author} {\bibfnamefont {W.~H.}\ \bibnamefont
  {Press}}, \bibinfo {author} {\bibfnamefont {S.~A.}\ \bibnamefont
  {Teukolsky}}, \bibinfo {author} {\bibfnamefont {W.~T.}\ \bibnamefont
  {Vetterling}}, \ and\ \bibinfo {author} {\bibfnamefont {B.~P.}\ \bibnamefont
  {Flannery}},\ }\href@noop {} {\emph {\bibinfo {title} {Numerical Recipes 3rd
  Edition: The Art of Scientific Computing}}}\ (\bibinfo  {publisher}
  {Cambridge University Press},\ \bibinfo {year} {2007})\ Chap.~\bibinfo
  {chapter} {20}\BibitemShut {NoStop}%
\bibitem [{\citenamefont {Weitz}\ \emph {et~al.}(2016)\citenamefont {Weitz},
  \citenamefont {Eksin}, \citenamefont {Paarporn}, \citenamefont {Brown},\ and\
  \citenamefont {Ratcliff}}]{WeitzE7518}%
  \BibitemOpen
  \bibfield  {author} {\bibinfo {author} {\bibfnamefont {J.~S.}\ \bibnamefont
  {Weitz}}, \bibinfo {author} {\bibfnamefont {C.}~\bibnamefont {Eksin}},
  \bibinfo {author} {\bibfnamefont {K.}~\bibnamefont {Paarporn}}, \bibinfo
  {author} {\bibfnamefont {S.~P.}\ \bibnamefont {Brown}}, \ and\ \bibinfo
  {author} {\bibfnamefont {W.~C.}\ \bibnamefont {Ratcliff}},\ }\href {\doibase
  10.1073/pnas.1604096113} {\bibfield  {journal} {\bibinfo  {journal}
  {Proceedings of the National Academy of Sciences}\ }\textbf {\bibinfo
  {volume} {113}},\ \bibinfo {pages} {E7518} (\bibinfo {year}
  {2016})}\BibitemShut {NoStop}%
\end{thebibliography}%
\end{document}


\pagenumbering{arabic}
\onecolumngrid
\begin{center}
\textbf{\textbf Supplementary information for the manuscript \\ ``Spatial interactions and oscillatory tragedies of the commons''}

\author{Yu-Hui Lin}
\affiliation{School of Physics, Georgia Institute of Technology, Atlanta, GA,
USA}
\author{Joshua S. Weitz}
\affiliation{School of Physics, Georgia Institute of Technology, Atlanta, GA,
USA}
\affiliation{School of Biological Sciences, Georgia Institute of Technology,
Atlanta, GA, USA}
\date{\today}
\maketitle
\end{center}

\onecolumngrid
\section{Derivation of Individual Game Rules Recovering Replicator Dynamics}
\label{sup_sec:master_equations}
Here we derive the individual based game rules that are able to recover replicator dynamics 
\begin{equation}
    x' = x\left(1-x\right)\left(r_C(x,n) - r_D(x,n)\right),
\end{equation}
in the continuous limit. The time is scaled by parameter $\epsilon$ for more compact expression. 
The normalized time $\frac{t}{\epsilon}$ will be denoted as $\tau$ hereafter. 
We express the time derivation with respect to normalized time with the prime symbol. 
We postulate that individuals undergo a birth-death process:
 a focal player that wins against its opponent in a game can reproduce a new
individual with the same strategy and replace (A) another random player in the 
system or (B) the opponent. Here we show the derivation process of the scenario where a game involves three players (A), 
which recapitulates the form of replicator dynamics first. The derivation following similar process but 
with only 2 individuals involved in a game will be briefly introduced after derivation of (A) to show 
how it fails.

\subsection{Individual-based rules involving 3 players in a game}
\label{sup_subsec:IBM3}
We write the events in a chemical reaction form with reaction rates
\(k_i\). The total number of players in the system is constant as a result of
the birth-death process, $N = n_C + n_D$, where $n_C$ and $n_D$ stand for 
number of cooperators and defectors respectively. 
With \(C\) denoting a cooperator and \(D\) denoting a
defector, the possible combinations of players involved in a game are
\begin{equation}\label{sup_eq:reactions}
	\begin{aligned}
		&\overbrace{C}^{\text{focal}} &+
		&\overbrace{C}^{\text{opponent}}\rule{-12pt}{0ex}&+
		&\overbrace{D}^{\substack{\text{random} \\ \text{player}}}\rule{-12pt}{0ex}
		&\xrightarrow{k_1}
		&\overbrace{C}^{\text{focal}} &+ 
		&\overbrace{C}^{\text{opponent}}\rule{-12pt}{0ex} &+
		&\overbrace{C}^{\text{replaced}} \\
		&\rule{7pt}{0ex}C &+ &\rule{13pt}{0ex}D &+ &\rule{12pt}{0ex}D &\xrightarrow{k_2} &\rule{7pt}{0ex}C &+ &\rule{13pt}{0ex}D &+ &\rule{12pt}{0ex}C \\
		&\rule{7pt}{0ex}D &+ &\rule{13pt}{0ex}C &+ &\rule{12pt}{0ex}C &\xrightarrow{k_3} &\rule{7pt}{0ex}D &+ &\rule{13pt}{0ex}C &+ &\rule{12pt}{0ex}D \\
		&\rule{7pt}{0ex}D &+ &\rule{13pt}{0ex}D &+ &\rule{12pt}{0ex}C &\xrightarrow{k_4} &\rule{7pt}{0ex}D &+ &\rule{13pt}{0ex}D &+ &\rule{12pt}{0ex}D \\
	\end{aligned}
\end{equation}
The transition rates of Eq.~\ref{sup_eq:reactions} are
\begin{equation}\label{sup_eq:trans_rates}
\begin{split}
\mathbb{T}(n_C+1 | n_C) &= k_1 \cdot n_C \cdot \overbrace{
\frac{n_C-1}{N-1}}^{\substack{\text{randomly choosing} \\ \text{an opponent C}}}\cdot\overbrace{ \frac{n_D}{N-1} }^{\substack{\text{a random D to} \\ \text{be
replaced}}}
+ k_2 \cdot n_C \cdot \overbrace{ \frac{n_D}{N-1}}^{\substack{\text{randomly
choosing} \\ \text{an opponent D}}}
\cdot\overbrace{ \frac{n_D}{N-1} }^{\substack{\text{a random D to} \\ \text{be
replaced}}} \\
		\\
\mathbb{T}(n_C-1 | n_C) &= k_3 \cdot n_D \cdot \overbrace{
\frac{n_C}{N-1}}^{\substack{\text{randomly choosing} \\ \text{an opponent C}}}
\cdot\overbrace{ \frac{n_C}{N-1} }^{\substack{\text{a random C to} \\ \text{be
replaced}}}
+ k_4 \cdot n_D \cdot \overbrace{ \frac{n_D-1}{N-1}}^{\substack{\text{randomly
choosing} \\ \text{an opponent D}}}
\cdot\overbrace{ \frac{n_C}{N-1} }^{\substack{\text{a random C to} \\ \text{be
replaced}}}.
\end{split}
\end{equation}
The terms in the above equations describe the change in number of two types of
players in a probabilistic perspective: there are \(n_C \) (\(n_D \))
cooperators (defectors) in the system, each of which has probability to pick
another player, either a cooperator (\(\frac{n_C - 1}{N-1} \)) or a defector
(\(\frac{n_D- 1}{N-1} \)), as an opponent. At each time increment, the focal
player can produce a new individual with reaction rate \(k_i \Delta \tau\). All
players other than the focal player in the system has equal chance to be
replaced by the newborn individual. The individual to be replaced is
a cooperator (defector) with probability $\frac{N_c - 1}{N-1}$ ($\frac{N_D - 1}{N-1}$).
Combining the transition probability of all the possible events listed above,
 the master equation that governs the change in $n_C$ is accordingly
\begin{equation}\label{sup_eq:master}
\begin{split}
\mathbb{P}\left(n_C, \tau+\Delta \tau \right) &= \mathbb{P}\left(n_C, \tau \right) + \mathbb{T}(n_C | n_C - 1 )\mathbb{P}(n_C- 1, \tau) \Delta \tau+ \mathbb{T}(n_C | n_C + 1 )\mathbb{P}(n_C + 1, \tau) \Delta \tau\\
&\ \ \ - \mathbb{T}(n_C+1 | n_C)\mathbb{P}(n_C, \tau) \Delta \tau- \mathbb{T}(n_C-1 | n_C)\mathbb{P}(n_C, \tau)\Delta \tau +
\mathcal{O}(\Delta \tau^2).
\end{split}
\end{equation}
Rearranging the master equation Eq.~\ref{sup_eq:master}, dividing both side by \(\Delta \tau \) 
and taking \(\Delta \tau \) to be infinitesimal yields
\begin{equation}\label{sup_eq:master2}
\begin{aligned}
\mathbb{P}'\left( n_C, \tau \right) = \mathbb{T}(n_C | n_C - 1 )\mathbb{P}(n_C - 1, \tau) + \mathbb{T}(n_C | n_C + 1 )\mathbb{P}(n_C+ 1, \tau) \\
	- \mathbb{T}(n_C+1 | n_C)\mathbb{P}(n_C, \tau) - \mathbb{T}(n_C-1 | n_C)\mathbb{P}(n_C, \tau).
\end{aligned}
\end{equation}
To examine if the master equation derived from the individual-based events 
recovers replicator dynamics in the continuous limit, we need to derive the time
differential equation of the expected value of $n_C$ from the master equation Eq.~\ref{sup_eq:master2}. 
To proceed, we denote the expected value of a variable with angular brackets. 
The definition of the expected value of a random variable \(X\) is
\begin{equation}
	\langle X (\tau) \rangle = \sum_{X} X\mathbb{P}(X, \tau).
\end{equation}
Next, we multiply each term in the master equation Eq.~\ref{sup_eq:master2} with \(n_C\) and sum over
 all possible values of \(n_C\) to obtain the time differential equation 
 describing the dynamics of \(\langle n_C \rangle\).
\begin{equation}\label{sup_eq:summation}
\begin{aligned}
\sum_{n_C=0}^{N} n_C \mathbb{P}' \left( n_C \right) &= \sum_{n_C=1}^{N} n_C \mathbb{T}(n_C |
n_C - 1 )\mathbb{P}(n_C - 1) + \sum_{n_C=0}^{N-1} n_C \mathbb{T}(n_C | n_C + 1 )\mathbb{P}(n_C + 1) \\
&\ \ \ \ \ \ \ \ \ \ \ \ - \sum_{n_C=0}^{N} n_C \mathbb{T}(n_C+1 | n_C)\mathbb{P}(n_C) -
\sum_{n_C=0}^{N} n_C \mathbb{T}(n_C-1 | n_C)\mathbb{P}(n_C) \\
&= \sum_{n_C=0}^{N} (n_C+1)\mathbb{T}(n_C+1 | n_C)\mathbb{P}(n_C) + \sum_{n_C=0}^{N}
(n_C-1)\mathbb{T}(n_C-1 | n_C)\mathbb{P}(n_C)\\
&\ \ \ \ \ \ \ \ \ \ \ \ - \sum_{n_C=0}^{N} n_C \mathbb{T}(n_C+1 | n_C)\mathbb{P}(n_C) -
\sum_{n_C=0}^{N} n_C \mathbb{T}(n_C-1 | n_C)\mathbb{P}(n_C) \\
&=\sum_{n_C=0}^{N} \mathbb{T}(n_C+1 | n_C)\mathbb{P}(n_C) - \sum_{n_C=0}^{N} n_C \mathbb{T}(n_C-1 |
n_C)\mathbb{P}(n_C).
\end{aligned}
\end{equation}
Plugging the expressions of transition rates in eq. Eq.~\ref{sup_eq:trans_rates} into
 the last equation Eq.~\ref{sup_eq:summation} leads to
\begin{equation}\label{sup_eq:avg_dot}
\begin{aligned}
\langle n_C\rangle' &= \langle \mathbb{T} \left( n_C+1 | n_C\right)\rangle - \langle \mathbb{T} \left( n_C-1 | n_C \right) \rangle \\
&\approx k_1\cdot \langle n_C \cdot \frac{n_C}{N} \cdot \frac{n_D}{N} \rangle +
k_2\cdot \langle n_C \cdot \frac{n_D}{N} \cdot \frac{n_D}{N} \rangle
- k_3 \cdot \langle n_D \cdot \frac{n_C}{N} \cdot \frac{n_C}{N} \rangle -
k_4\cdot \langle n_D \cdot \frac{n_C}{N} \cdot \frac{n_D}{N} \rangle.
\end{aligned}
\end{equation}
when \(N, n_C, n_D >> 1\).
As stated earlier, $N = n_C + n_D $ is a constant in the system. 
We then define \(x \equiv
\frac{n_C}{N}\), and thus \(\frac{n_D}{N} = 1-x\). Dividing both sides of Eq.~\ref{sup_eq:avg_dot} by \(N\)
, we find
\begin{equation}
\langle x\rangle' = k_1\cdot \langle x^2(1-x)\rangle + k_2\cdot \langle
x(1-x)^2\rangle - k_3 \cdot \langle x^2(1-x)\rangle - k_4\cdot \langle
x(1-x)^2\rangle.
\end{equation}
For sufficiently large populations, fluctuations around the average value are expected
to be sufficiently small, and by ignoring correlatinos, we
approximate the expected values with actual values and omit the angular brackets. 
Rearranging Eq.~\ref{sup_eq:avg_dot2} and omitting
angular brackets, we get
\begin{equation}
\begin{aligned}\label{sup_eq:avg_dot2}
	x' &= x(1-x) \left[ (k_1 x + k_2 (1-x)) - (k_3 x + k_4 (1-x))\right],
\end{aligned}
\end{equation}
which recovers replicator dynamics with growth rates
\[
	\begin{bmatrix}
		r_C \\
		r_D
	\end{bmatrix}
	=
	\begin{bmatrix}
		k_1 & k_2 \\
		k_3 & k_4
	\end{bmatrix}
	\begin{bmatrix}
		x \\
		1-x
	\end{bmatrix}
	=
	\begin{bmatrix}
		k_1 x + k_2 (1-x) \\
		k_3 x + k_4 (1-x)
	\end{bmatrix}.
\]
Compared with Eq. 2  in the main text, it is apparent that
the values of \(k_i\)'s should be
\[ 
\begin{cases}
	k_1 &= R(n) \\
	k_2 &= S(n) \\
	k_3 &= T(n) \\
	k_4 &= P(n) \\
\end{cases}.
\]
which correspond to the payoffs of focal players.
\subsection{Individual-based rules involving 2 players in a game}
Interactions between individuals in a chemical equation form with reaction rates $k_i$ are
\begin{equation}\label{sup_eq:reactions_IBM2}
	\begin{aligned}
        \overbrace{C}^{\text{focal}} + \overbrace{D}^{\text{opponent}}  &\xrightarrow{k_1'}
        \overbrace{C}^{\text{focal}} + \overbrace{C}^{\substack{\text{opponet} \\ \text{replaced}}} \\
        D \ \ + \ \ \ C \ \ \ \ &\xrightarrow{k_2'} \ \ D \ +\ \  \ D 
	\end{aligned}
\end{equation}
Transition rates of Eq.~\ref{sup_eq:reactions_IBM2} are
\begin{equation}\label{sup_eq:trans_rates_IBM2}
    \begin{split}
    \mathbb{T}(n_C+1 | n_C) &= k_1' \cdot n_C \cdot \overbrace{
    \frac{n_D}{N-1}}^{\substack{\text{randomly choosing} \\ \text{an opponent C}}}
    		\\
    \mathbb{T}(n_C-1 | n_C) &= k_2' \cdot n_D \cdot \overbrace{
    \frac{n_C}{N-1}}^{\substack{\text{randomly choosing} \\ \text{an opponent C}}}.
     \end{split}
\end{equation}
Following the same derivation from Eq.~\ref{sup_eq:master2} to Eq.~\ref{sup_eq:summation} and 
plugging the expressions of transition rates in Eq.~\ref{sup_eq:trans_rates_IBM2}, 
\begin{equation}\label{sup_eq:avg_dot_IBM2}
    \begin{aligned}
        \langle n_C\rangle' &= \langle \mathbb{T} \left( n_C+1 | n_C\right)\rangle - \langle \mathbb{T} \left( n_C-1 | n_C \right) \rangle \\
        &\approx k_1'\cdot \langle n_C \cdot \frac{n_D}{N}  \rangle -
        k_2'\cdot \langle n_D \cdot \frac{n_C}{N} \rangle,
    \end{aligned}
\end{equation}
when $N, n_C, c_D >> 1$.
As before, $N = n_C + n_D $ is a constant in the system. We  
define \(x \equiv \frac{n_C}{N}\) and thus \(\frac{n_D}{N} = 1-x\). Dividing both sides of 
Eq.~\ref{sup_eq:avg_dot_IBM2} gives
\begin{equation}
\langle x\rangle' = k_1'\cdot \langle x(1-x)\rangle - 
k_2'\cdot \langle x(1-x)\rangle. 
\end{equation}
For large populations, the fluctuations around expected values are negligible, and
by assuming the absence of higher-order correlation, we find
\begin{equation}
x = (k_1' - k_2')x(1-x).
\end{equation}
This mean field dynamics in the IBM2 model is quadratic in $x$ assuming $k_i$'s do not dependent on information
other than payoffs, rather than being cubic in $x$ as expected in the replicator dynamics for the IBM3 model.
\section{Simulation Details}
\subsection{Individual-based simulation}
To simulate the coevolutionary dynamics in an individual-based, game-theoretic perspective, 
we adopt the individual-based game rules derived in section~\ref{sup_subsec:IBM3}. We simulate $N$ games 
sequencially to allow equal chance of each individual to be involved in a game within a time step $\Delta t$. 
At the beginning of each time step, each player is labelled by an integer ranging from $1$ to $N$.
3 randomly permuted number series ranging from $1$ to $N$ are
 generated. The $i-$th elements in the three random number series represent 
 the index of the focal player, the opponent,  the individual to be replaced 
 in the $i-$th game respectively.   
 A random number $r \in \left[0,1\right)$ is generated for each game, 
 and the birth-death process happens if $r > k_i \Delta \tau = k_i \frac{\Delta t}{\epsilon}$. 
 $k_i$ is the reaction rate and depends on the strategies of the focal player and the 
 opponent. 
 We complete a time step by updating environment $n$ with Euler method, 
 \[
 \Delta n = \dot{n} \Delta t = n(1-n)(\theta x_{new}(t) - (1-x_{new}(t))) \Delta t,
 \] 
 after $N$ games. The subscript $new$ denotes that the value of $x$ in $\dot{n}$ 
 is taken to be the one after $N$ games, rather than the value at the beginning of the simulation time step.
A simulation proceeds to the next time step after both $x$ and $n$ are updated, and stops when 
total number of time steps reaches an assigned simulation horizon.
\subsection{Spatially-explicit simulation}
We perform spatially-explicit simulations on a 2D $L\times L$ lattice. 
Each lattice site can only  be occupied by a single player, either a cooperator or a defector.
The rules for updating individual strategy in spatially-explicit simulations are similar to that in non-spatial simulations.
Instead of generating three random number series, only a randomly permuted number series is generated to represent the index of the focal player in each game. 
The opponent player and the individual to be replaced are then chosen in the von Neumann neighborhood of the focal player.

After $N$ sequential games, the spatial profile of the environment is updated in two steps. 
\begin{subequations}
    \begin{align}
	n_{i,j, s}(t) = & n_{i,j}(t) + n_{i,j}(t)(1-n_{ij}(t))\left(\theta x_{i,j,new} - (1-x_{i,j,new}) \right) \label{sup_eq:dn_strategy} \\
	n_{i,j, f}(t+\Delta t) = &
	\begin{cases}
		\begin{array}{l} 
			D_n \frac{\Delta t}{\Delta x^2}(n_{i-1,j, s}(t)-2n_{i,j, s}(t)+n_{i+1,j, s}(t) \\
				\ \ \ \ \ + n_{i,j-1, s}(t)-2n_{i,j, s}(t)+n_{i,j+1, new}(t))  \label{sup_eq:dn_diff}
		  \end{array}, &\text{for } D_n=0 \text{ or  } 1 \\
		\frac{\Sigma_{i,j}n_{i,j,s}(t)}{L \cdot L}, &\text{for } D_n=\infty
	\end{cases}.
    \end{align}
\end{subequations}
The first step (Eq.~\ref{sup_eq:dn_strategy}) accounts for change in $n$ due to individual 
strategies adopted in previous games in the same time step.
The second step accounts for diffusion was calculated with standard explicit 
forward-time centered-space method (Eq.~\ref{sup_eq:dn_diff}).
The subscripts $i,j$ denote the value at coordinate $(i, j)$. 
The subscript $new$ in $x_{new}$ again denotes the updated quantity after $N$ games.
$s$ in $n_{i,j,s}$  means the value of $n$ is only affected by the individual strategies 
but has not yet accounted for the effect of diffusion. $f$ in $n_{i,j,f}$ indicates this is the final value for $n$ after a 
time step $\Delta t$. As before, a spatial simulation proceeds to the next simulation step after both $x$ and $n$ are updated
and stops at some pre-assigned simulation horizon.
\subsection{Simulation parameters}

We choose $\Delta x = 1$ and $\Delta t = 0.05$ to ensure the stability of 2D diffusion algorithm. Stability is ensured when
$D_n \frac{\Delta t}{\Delta x^2} \le \frac{1}{4} $\cite{press2007numerical}. We use periodic boundary conditions for spatial simulations. 
The scaling value $\epsilon = 0.5$ was further chosen so that $\Delta \tau = \Delta t / \epsilon = 0.1$.
This assures the largest transition probability of a birth-death process $k_i \Delta \tau$ will be no larger than $1$ given 
the values of payoff matrix chosen in all the simulations.
The model parameters are in arbitrary simulation units,
and the values other than $A_0$ are shown in Table~\ref{sup_table:parameters}.

\begin{figure}[H]
	\centering
	\includegraphics[width=.5\textwidth]{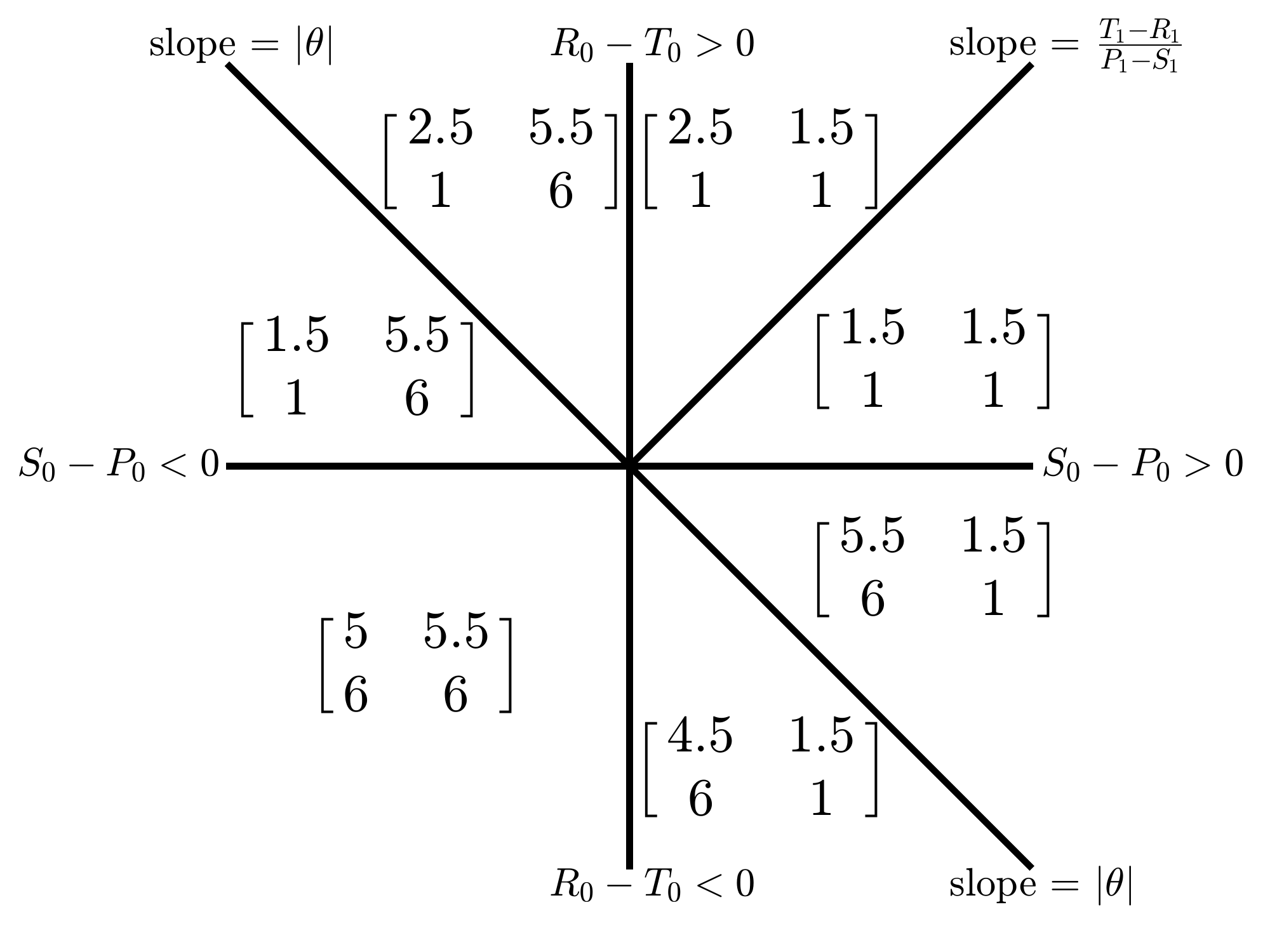}
	\caption{The values of matrix $A_0 = \big[ \begin{smallmatrix} R_0& S_0\\ T_0&P_0 \end{smallmatrix} \big]$ in each section on the parameter space of FIG. 1 in the main text.}
	\label{sup_fig:A0_fig1}
\end{figure}

\begin{figure}[H]
        \centering
	\includegraphics[width=.5\textwidth]{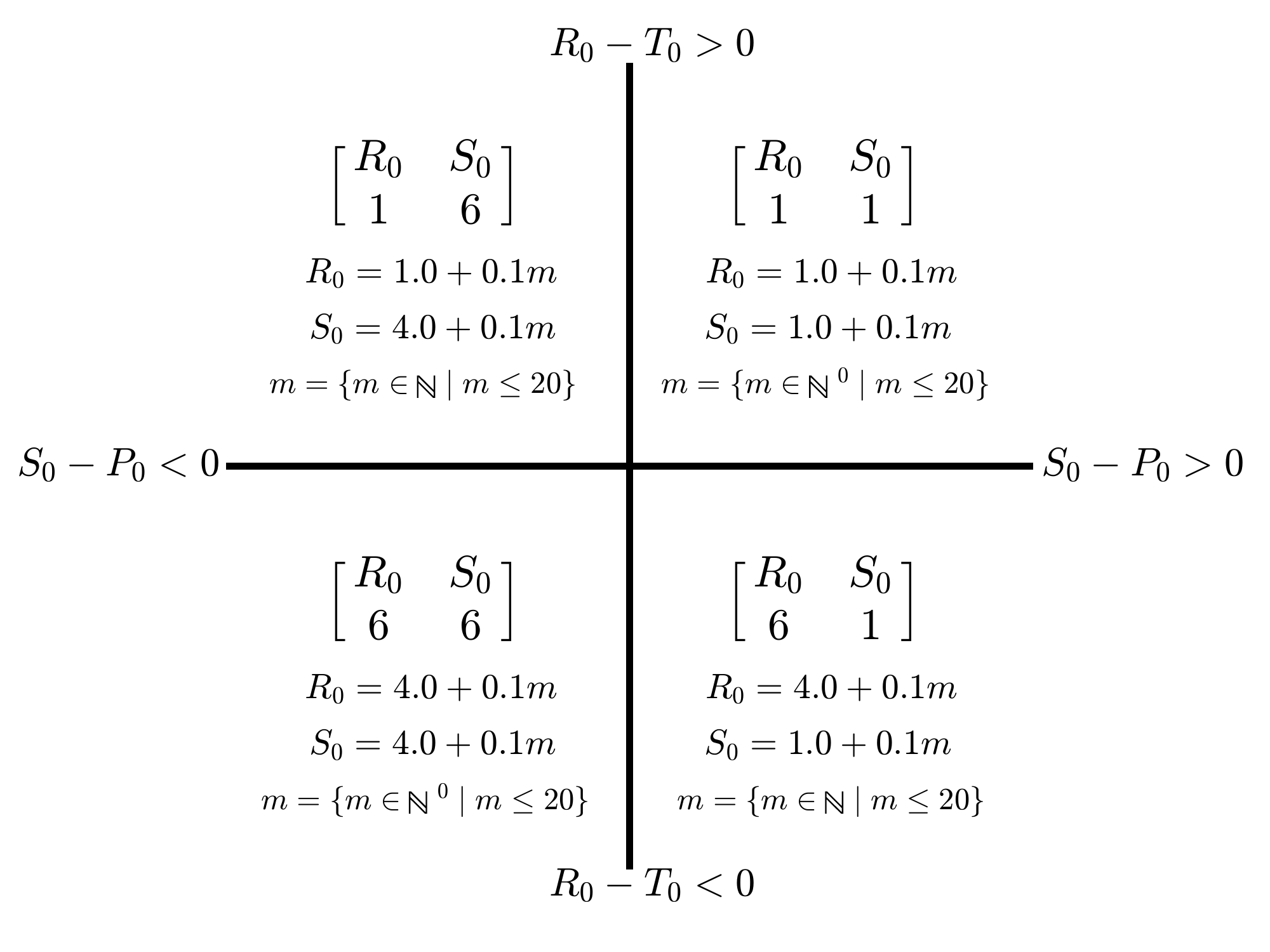}
	\caption{The values of matrix $A_0 = \big[ \begin{smallmatrix} R_0& S_0\\ T_0&P_0 \end{smallmatrix} \big]$ in each section on the parameter space of FIG. 2 in the main text.}
	\label{sup_fig:A0_fig2}
\end{figure}

The parameters of simulations in FIG. 1 are shown in Table~\ref{sup_table:parameters} and FIG.~\ref{sup_fig:A0_fig1}.
In FIG. 1, each plot for stochastic individual-based model dynamics are averaged over 100 replicates 
except for oscillatory dynamics, given phase differences that can arise due to demographic noise.
The parameters except for $D_n$ of simulations in FIG. 2 are shown in Table~\ref{sup_table:parameters} and FIG.~\ref{sup_fig:A0_fig2}. 
$D_n$ is specified on each panel of FIG. 2 in the main text. The axes of heat maps ranging from -2 to 2, 
linearly increase with difference 0.1 on both the horizontal axis, $S_0-P_0$ and the vertical axis, $R_0 - T_0$. 
There are thus $41 \cdot 41 = 1681$ grids on each heat map.  
The value of each grid on the heat maps is the average over 20 replicates, so there are information of $41 \cdot 41 \cdot 20 \cdot 4 = 134480$
simulations in FIG. 2. 
The parameters except for $D_n$ of simulations in FIG. 3 are shown in Table~\ref{sup_table:parameters} and $A_0 = \big[ \begin{smallmatrix} 2.5 & 5.5\\ 1&6 \end{smallmatrix} \big]$. 
$D_n$ is specified on each panel of FIG. 3 in the main text. 

\begin{table}[H]
\centering
\caption{Model parameter values in simulation units.}
\label{sup_table:parameters}
\begin{tabular}{cll}
\hline 
\bf{Variable}                 & \multicolumn{1}{c}{\bf{Value}}                      \\
\hline
$A_1$       & $\big[ \begin{smallmatrix} 3&0\\ 5&1 \end{smallmatrix} \big]$                       \\[5pt]
$A_0$         & see below for each figure \\[5pt]
$N$  & 10000                        \\
$\epsilon$  & 0.5                        \\
$\theta$         & 2                          \\
$D_n$                & [0, 1, $\infty$]     	\\
$L$                & 100     	\\
$(x_0, n_0)$                & $(0.3, 0.7)$     	\\
                  
\end{tabular}
\end{table}

\subsection{Classification Criteria for TOC and averted dynamics}
The classification criteria for characterizing temporal dynamics beyond the transient phase of dynamics into TOC or non-TOC is as follows:
\begin{itemize}
	\item Find the difference between consecutive peaks and valleys in the environment, $\delta n$, and in the fraction of cooperators, $\delta x$.
	\begin{itemize}
		\item if the maximum $\delta n$ or $\delta x$ is larger than the threshold $1-2\delta$, it indicates the dynamics is an oscillating TOC because of the large variation in magnitude. $\delta$ is a smaller number set to be $0.01$.
		\item else, we take the mean of the later $20\%$ time series of $x(t)$ and $n(t)$, i.e., $\bar{x}$ and $\bar{n}$ respectively.  The classification into a TOC is as follows:
			\begin{itemize}
				\item if $\bar{n} < \delta$, it is a TOC.
				\item else, a TOC is averted.
			\end{itemize}
		
	\end{itemize}
\end{itemize}

To separate quasi-periodic dynamics from an oscillating TOC, we categorize a dynamics as 
an o-TOC if the amplitude is large enough (maximum $\delta n > 1-2\delta$ or $\delta x > 1-2\delta$).
This decision stump allows us to classify dynamics with small to moderate oscillations as averted. 
After separating o-TOC from other oscillating dynamics, a trajectory is classified
as a TOC if the environment is drained at the later part of the simulation, 
namely $\bar{n} < \delta$ regardless what the value of $\bar{x}$ is.
\section{Spatial-Temporal Dynamics Simulation Results}
\begin{figure}[H]
	\centering
	\begin{subfigure}{0.9\textwidth} 
		\caption{}
		\includegraphics[width=\textwidth]{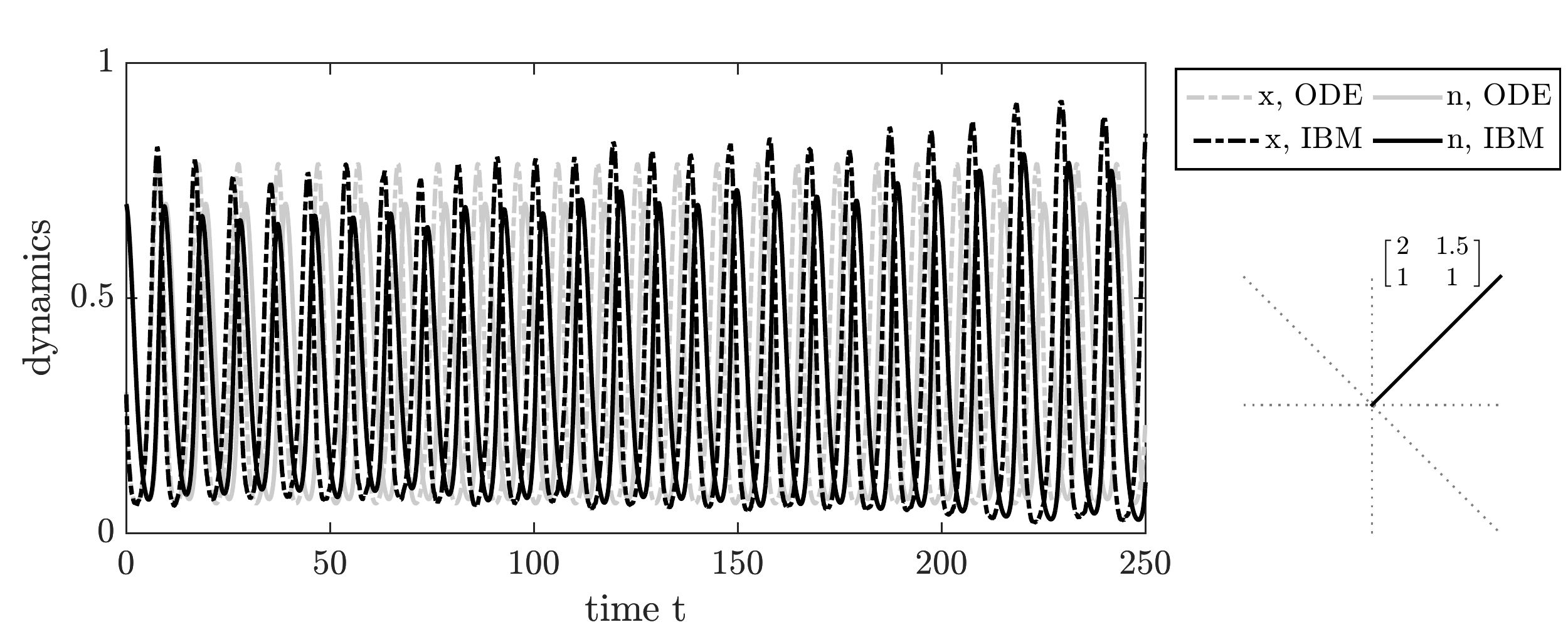}
	\end{subfigure}
	\vspace{1em} 
	\begin{subfigure}{0.9\textwidth} 
		\caption{} 
		\includegraphics[width=\textwidth]{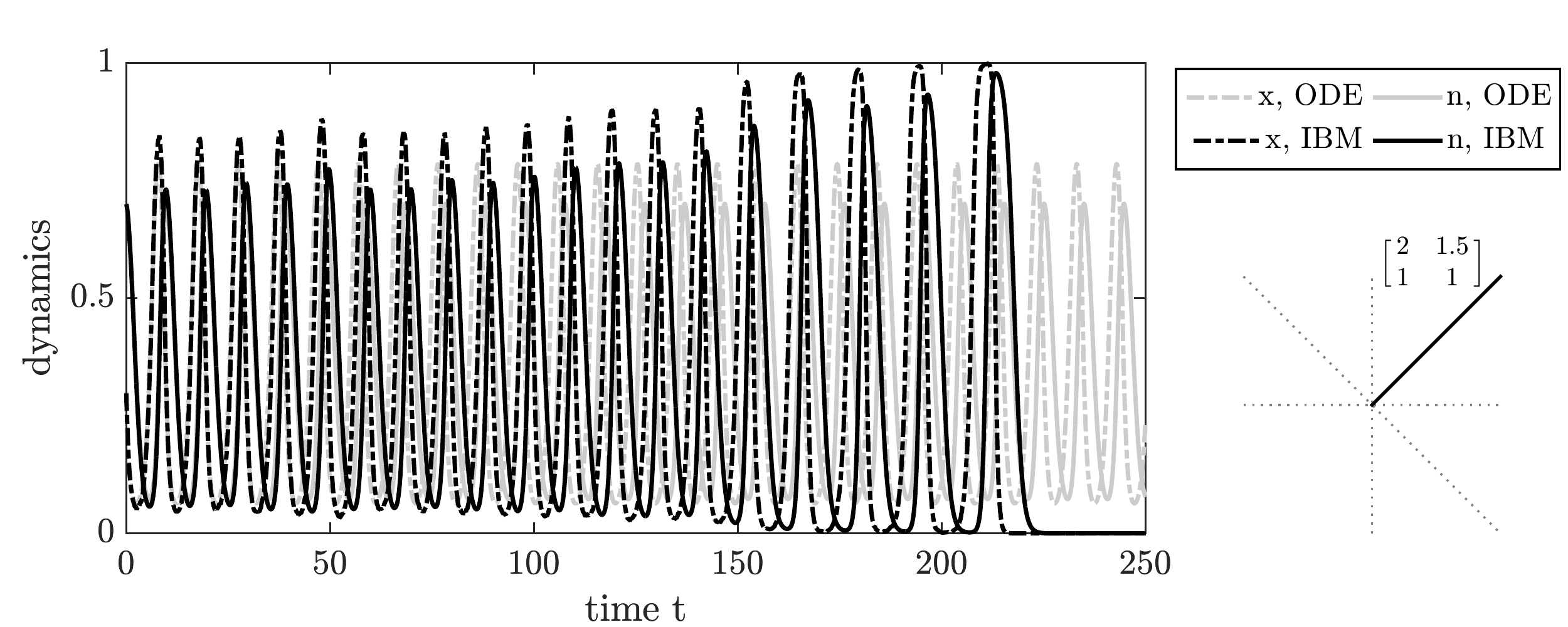}
		\label{sup_fig:demographic_b}
	\end{subfigure}
\caption{Demographic noise can alter the dynamical behaviors of the coevolutionary system. 
(a) Persistent oscillations with varying amplitudes with IBM (black lines), (b) If the amplitude of oscillations grows larger, it is possible for the
system to approach one of the absorbing states $(x, n) = (0, 0)$, $(0, 1)$, $(1, 0)$, $(1, 1)$.
The values of $A_0$ in this Figure are
 $\big[ \begin{smallmatrix} 2& 1.5\\ 1&1 \end{smallmatrix} \big]$.}
\label{sup_fig:demographic}
\end{figure}

FIG.~\ref{sup_fig:demographic} shows how demographic noise in non-spatial 
individual based models (IBM) can change the system dynamics. 
The analysis of the ODE system (Eq. 2 and Eq. 3 in the main text) predicts 
neutral orbits when $(R_0-T_0)/(S_0-P_0) = \theta$ (gray lines) \cite{WeitzE7518}, 
but the non-spatial IBM simulations reveal that the perfect orbits are 
impossible due to stochasticity.
We can see in FIG.~\ref{sup_fig:demographic}
that demographic noise modulates the amplitudes of oscillations 
substantially. The increased amplitudes can drive the system
toward a boundary and
may even lead to a TOC (FIG.~\ref{sup_fig:demographic_b}).
Oscillations may also be induced by strong coupling between 
local sites mediated by diffusing environment. 
As shown in FIG.~\ref{sup_fig:Dn_inf_oscillation}, oscillations
are present 
across a large range of the $(S_0-P_0)-(R_0-T_0)$ parameter space.
 Some of the oscillations may lead to a TOC (FIG.~\ref{sup_fig:Dn_inf_oscillation_a}),
while others may have various amplitudes but the population persists over
multiple runs, in these cases until the end of the simulation
 (FIG.~\ref{sup_fig:Dn_inf_oscillation_b} - FIG.~\ref{sup_fig:Dn_inf_oscillation_d}).
Given the absorbing conditions of the stochastic model, we cannot
guarantee infinite-time persistence of oscillations in any finite 
simulation.
\begin{figure}[H]
	\centering
	\begin{subfigure}{.9\textwidth} 
		\caption{}
		\includegraphics[width=\textwidth]{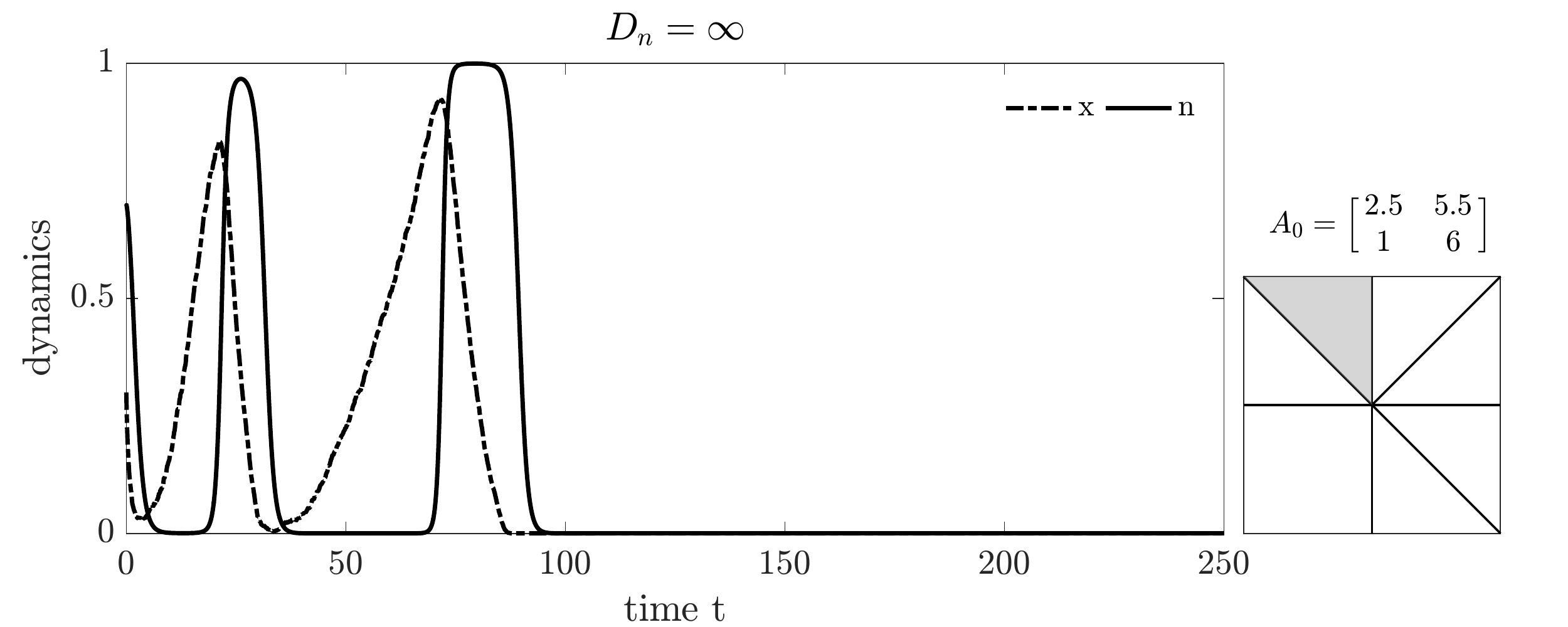}
		 \label{sup_fig:Dn_inf_oscillation_a}
	\end{subfigure}
	\vspace{1em} 
	\begin{subfigure}{.9\textwidth} 
		\caption{} 
		\includegraphics[width=\textwidth]{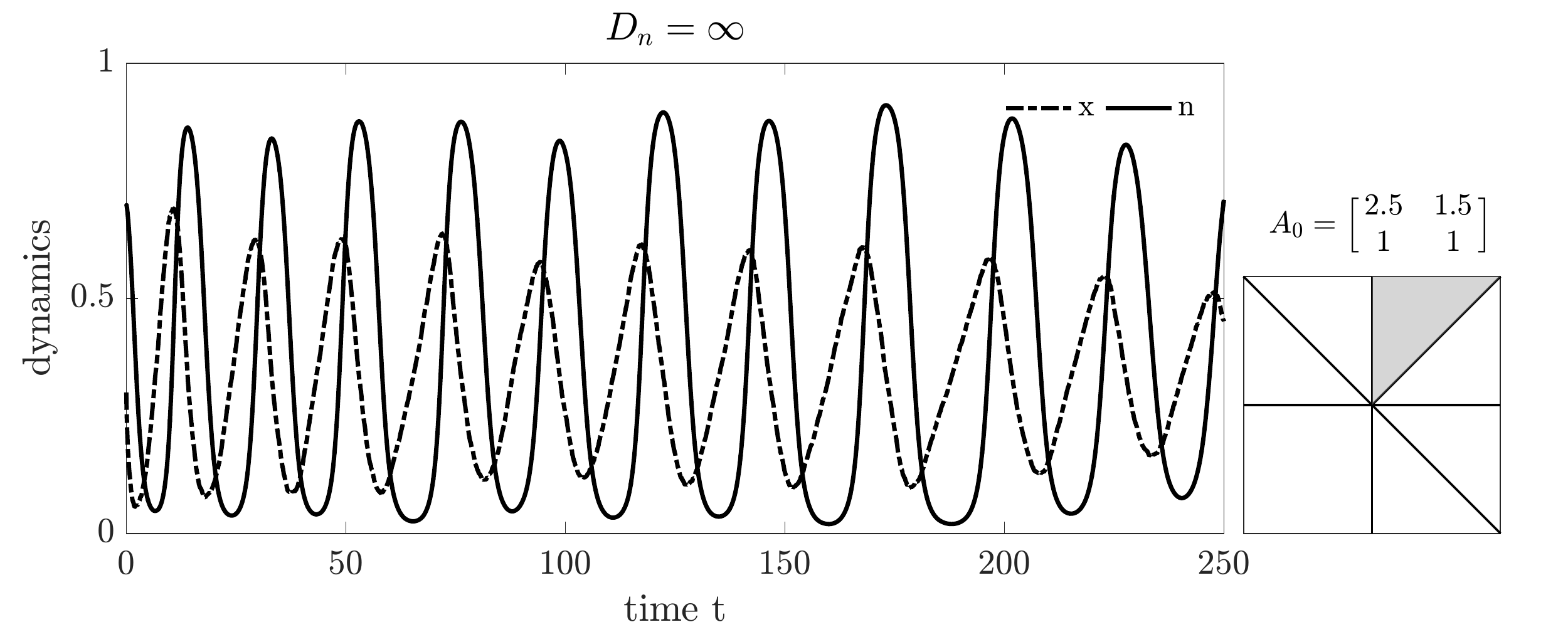}
		\label{sup_fig:Dn_inf_oscillation_b}
	\end{subfigure}
	
	\begin{subfigure}{.9\textwidth} 
		\caption{}
		\includegraphics[width=\textwidth]{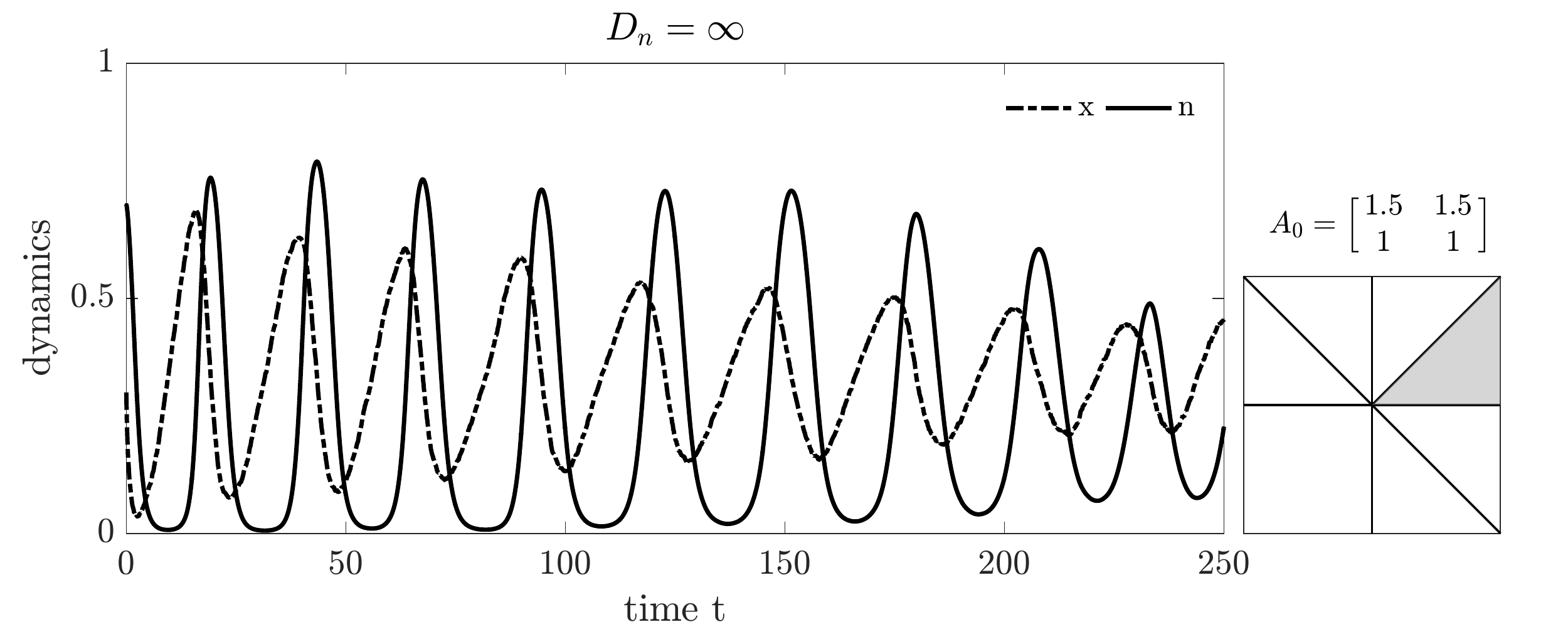}
		\label{sup_fig:Dn_inf_oscillation_c}
	\end{subfigure}
\end{figure}
\begin{figure}\ContinuedFloat
	\vspace{1em} 
	\begin{subfigure}{.9\textwidth} 
		\caption{} 
		\includegraphics[width=\textwidth]{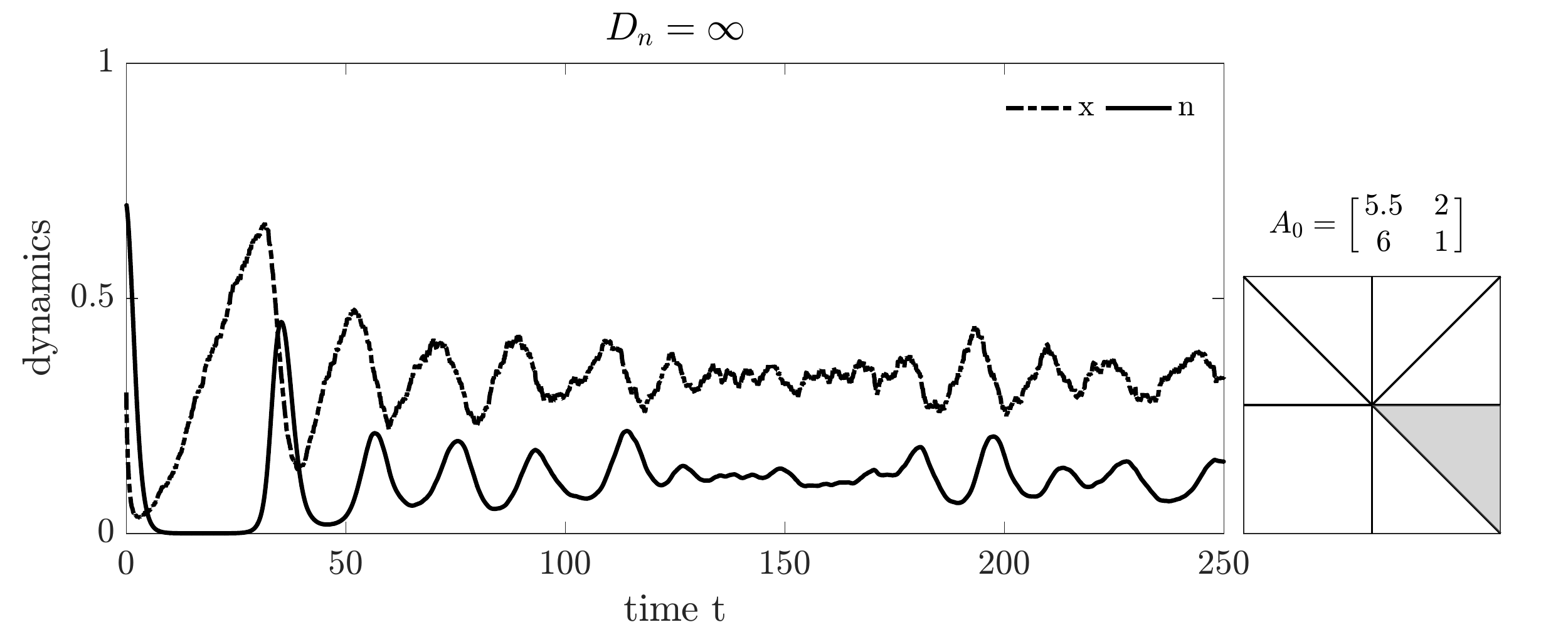}
		\label{sup_fig:Dn_inf_oscillation_d}
	\end{subfigure}
\caption{The oscillating dynamics are common with $D_n = \infty$ in a wide range of $A_0$. These
four plots shows the temporal dynamics of a replicate with a specific $A_0$, and where it falls on 
the $(S_0-P_0) - (R_0 - T_0)$ parameter space in the each inserted plot. The values of $A_0$'s are
 (a) $\big[ \begin{smallmatrix} 2.5& 5.5\\ 1&6 \end{smallmatrix} \big]$ , 
 (b) $\big[ \begin{smallmatrix} 2.5& 1.5\\ 1&1 \end{smallmatrix} \big]$,
 (c) $\big[ \begin{smallmatrix} 1.5& 1.5\\ 1&1 \end{smallmatrix} \big]$, and  
 (d) $\big[ \begin{smallmatrix} 5.5& 2\\ 6&1 \end{smallmatrix} \big]$, respectively.}
 \label{sup_fig:Dn_inf_oscillation}
\end{figure}\bibliography{refs}